\documentclass[manuscript]{emulateapj}

\def\Zsol{\hbox{Z$_{\odot}$}}
\def\Msol{\hbox{M$_{\odot}$}}

\usepackage{epsfig,natbib,url,graphicx}
\usepackage{color}
\usepackage{soul} 
\usepackage{fancyvrb}
\usepackage{pifont}
\usepackage{longtable,enumitem}

\newcommand{\kms}{km\,s$^{-1}$}
\newcommand{\hi}{H\,{\sc i}}
\newcommand{\hii}{H~{\sc ii}}

\newcommand{\heii}{He~{\sc ii}}
\newcommand{\eld}{$N_{\rm e}$}

\newcommand{\elt}{$T_{\rm e}$}

\newcommand{\foiii}{[O~{\sc iii}]}

\newcommand{\foii}{[O~{\sc ii}]}
\newcommand{\fsii}{[S~{\sc ii}]}
\newcommand{\fsiii}{[S~{\sc iii}]}

\newcommand{\fnii}{[N~{\sc ii}]}

\newcommand{\oii}{O~{\sc ii}}

\newcommand{\ffev}{[Fe~{\sc v}]}

\newcommand{\ha}{H$\alpha$}
\newcommand{\hb}{H$\beta$}

\def\cm3{\hbox{cm$^{-3}$}}

\shorttitle{Imaging Ionisation and Metallicity in Mrk71}
\shortauthors{James et al.}

\begin{document}


\title{Resolving Ionisation and Metallicity on Parsec Scales\\ Across Mrk~71 with HST-WFC3 }
\author{Bethan L. James\altaffilmark{1}\thanks{E-mail:bjames@ast.cam.ac.uk}, Matthew Auger\altaffilmark{1}, Alessandra Aloisi\altaffilmark{2}, Daniela Calzetti\altaffilmark{3} and Lisa Kewley\altaffilmark{4,5}}
\affil{1. Institute of Astronomy, University of Cambridge, Madingley Road, Cambridge, CB3 0HA, UK}
\affil{2. Space Telescope Science Institute, 3700 San Martin Drive, Baltimore, MD 21218, USA}
\affil{3. Department of Astronomy, University of Massachusetts, Amherst, MA 01003, USA}
\affil{4. RSAA, Australian National University, Cotter Road, Weston Creek, ACT 2611, Australia}
\affil{5. Institute for Astronomy, 2680 Woodlawn Drive, University of Hawaii, Hilo, HI 96822, USA}




\begin{abstract}
Blue Compact Dwarf (BCD) Galaxies in the nearby Universe provide a means for studying feedback mechanisms and star-formation processes in low-metallicity environments in great detail.  Due to their vicinity, these local analogues to young galaxies are well suited for high-resolution studies that would be unfeasible for primordial galaxies in the high-redshift universe.  Here we present $HST$-WFC3 observations of one such BCD, Mrk~71, one of the most powerful local starbursts known, in the light of \foii, \heii, \hb, \foiii, \ha, and \fsii.  At $D\simeq3.44$~Mpc, this extensive suite of emission line images enables us to explore the chemical and physical conditions of Mrk~71 on $\sim2$~pc scales.  Using these high spatial-resolution observations, we use emission line diagnostics to distinguish ionisation mechanisms on a pixel-by-pixel basis and show that despite the previously reported hypersonic gas and super-bubble blow out, the gas in Mrk~71 is photoionised, with no sign of shock-excited emission.  Using strong-line metallicity diagnostics, we present the first `metallicity image' of a galaxy, revealing chemically inhomogeneity on scales of ~$<$50~pc.  We additionally demonstrate that while chemical structure can be lost at large spatial scales, metallicity-diagnostics  can break down on spatial scales smaller than a \hii\ region.  \heii\ emission line images are used to identify up to six Wolf-Rayet stars in Mrk~71, three of which lie on the edge of blow-out region.  This study not only demonstrates the benefits of high-resolution spatially-resolved observations in assessing the effects of feedback mechanisms, but also the limitations of fine spatial scales when employing emission-line diagnostics.  Both aspects are especially relevant as we enter the era of extremely large telescopes, when observing structure on $\sim10$~pc scales will no longer be limited to the local universe.
\end{abstract}

\section{Introduction}

Observations of Blue Compact Dwarf (BCD) Galaxies in the nearby Universe provide a means for studying chemical evolution and star formation processes in chemically un-evolved environments.  BCDs are thought to have experienced only very low-level star-formation in the past, but are currently undergoing bursts of star-formation in relatively pristine environments, ranging from 1/50--1/2 \Zsol\, \citep{Kunth:2000}, making them ideal analogues to young `building-block' galaxies in the high-$z$ primordial universe \citep{Searle:1972}.  

One of the most powerful aspects of this analogous population is the fine spatial scales on which we can explore them compared to high-$z$ galaxies - allowing us to gain insight into chemical and physical processes that occur on sub-kpc scales within metal-poor environments.  Spatially resolved observations of these systems are growing rapidly in both number and quality, due to increasing IFU (integral field unit) capabilities (with seeing-limited spatial resolutions of 0.2$''$) and the initiation of extensive volume-limited IFU surveys such as ATLAS~3D \citep{Cappellari:2011},  SAMI \citep{Bryant:2015} and MANGA \citep{Bundy:2015}.  Moreover, with the addition of adaptive-optics systems, IFU observations are now able to probe emission lines on ultra-fine spatial scales \citep[e.g. $\sim3$~pc per spaxel at $\sim$4~Mpc,][]{Muller:2014}.  As we enter 30--40~m telescope era, such scales will also be reachable in the more distant Universe, enabling us to resolve and explore structure in star-forming galaxies at the peak of cosmic star-formation  ($z=$2--3).  Until then, however, we continue to use high resolution observations of their nearby counterparts, BCDs, to further our understanding of the intricacies of galaxy formation and evolution, and prepare us for the types of structure that we will eventually probe at high-$z$.

Much effort is already underway to understand the ISM of galaxies on $\lesssim10$~pc scales. Hydrodynamical simulations at high spatial resolutions (i.e 1.5--12~pc) are being used to understand the triggering and evolution of star-formation \citep[e.g.][]{Teyssier:2010,Renaud:2014,Perret:2014}.  For example, by modelling a `realistic' ISM (i.e. inhomogeneous and thermally unstable), \citet{Teyssier:2010} find that starbursts are not primarily due to large-scale inflows of gas but rather gas fragmentation into massive and dense clouds.  However, many questions concerning the details of ISM structure of star-forming galaxies still remain open. 

Firstly, we must understand the sources of ionisation and the ionisation processes within BCDs. At present BCDs are modelled as giant \hii\ regions where photoionisation is the dominant ionisation mechanism influencing the interstellar medium (ISM).  However, the discovery of BCDs with significantly broad emission lines indicates that feedback mechanisms such as shocks, circumstellar emission, or even AGN may also play a significant role \citep[e.g.][]{Izotov:2007}.  Such feedback mechanisms are thought to have a strong influence on the evolution of dwarf galaxies \citep{Marconi:1994,Martin:1998,Ferrara:2000,Calzetti:2004}, although exact details of the process are still uncertain and depend on the nature and geometry of the galaxy \citep{DeYoung:1994}.  The energy injected into the ISM by supernova explosions can act as regulating mechanisms by heating and/or removing gas from the site of star-formation \citep{Kennicutt:1989,Heckman:1997,MacLow:1999}, and quenching the star-formation, whereas ionisation and shock fronts through the ISM may cause star-formation to propagate spatially \citep{Elmegreen:1977,McCray:1987}.  Processes like these, that suppress star-formation, have been suggested as solutions for the inefficiency of dwarf galaxies to form stars \citep{Guo:2010} and for the `missing satellites' problem, i.e. the discrepancy between observed and CDM-predicted numbers of dwarf galaxies \citep{Klypin:1999,Bovill:2009}.

Secondly, the question of chemical homogeneity of star-forming galaxies is still under debate.  Oxygen abundance maps of star-forming galaxies have shown them to be both chemically homogeneous \citep[e.g.,][]{James:2010,Kehrig:2008,Kehrig:2013,Perez-Montero:2011} and with structure \citep[e.g.,][]{Lopez-Sanchez:2011,James:2013a,James:2013b,Monreal-Ibero:2012,Sanchez-Almeida:2014b}, implying that feedback mechanisms can have a range of effects on chemical homogeneity, and also that a range of ISM mixing timescales exists within these galaxies on small spatial scales.  Such properties of \hii\ regions and starburst galaxies have been a controversial topic for many decades \citep{Kennicutt:2001, Kobulnicky:1996, Kobulnicky:1997} and are essential in constraining chemical evolution models. 

Thirdly, the effects of stellar feedback in low-metallicity environments are yet be constrained in detail. O-stars within BCDs can reach 10$^2$-$10^5$ in number \citep{Guseva:2000}.  Such a population has a profound influence on the dynamics of the gas, with stellar winds and supernovae  giving rise to outflows, superbubbles and shocks.   In such low metallicity environments, fast shocks combined with dense interstellar media have been theorised to explain the narrow high-ionization emission lines (e.g. \heii, \ffev), seen in the spectra of several BCDs with broad-component emission \citep{Thuan:2005}.  However, high densities are largely uncommon in BCDs, where densities typical of \hii\ regions ($\sim100$\,\cm3) are usually detected. WR stars (known to exist in more than 200 BCDs) can also produce characteristic broad emission in specific ions or `WR-features' (e.g. \heii\,$\lambda$4686) that originate in the envelopes of massive stars undergoing rapid mass loss.  Recent attempts to map galaxies in the light of \heii\,$\lambda$4686 have identified and mapped the WR stellar population \citep[e.g.][]{James:2009,Kehrig:2013,Kehrig:2015,Westmoquette:2013}, although we are yet to assess its role by spatially correlating the WR features to the shock and feedback morphologies on fine spatial scales.

The most robust way to investigate these processes in BCDs is to spatially resolve their emission lines and subsequent chemical and physical properties,  and to connect these to the resolved stellar populations. This can be achieved in high spatial resolution with the use of \textit{HST} narrow-band imaging. Previous work \citep[e.g.][]{Calzetti:1999, Calzetti:2004,Hong:2011} has shown that this method is highly efficient in investigating non-radiative ionisation processes in local star-forming galaxies. The angular resolution of \textit{HST} is crucial here, since shocks (one of the primary non-photoionised mechanisms expected) tend to create very thin ($<$10\,pc) and filamentary fronts, which cannot currently be detected with ground-based images, spectra or spatially resolved spectroscopy, as they become washed out by the `ocean' of photoionised gas emission \citep[e.g.][]{Calzetti:2004}. For example, the WFC3/UVIS pixel size of 0.04$''$ implies that structure down to $\sim$1\,pc can be resolved within nearby (D$\leq5$~Mpc) BCDs, allowing one to trace the morphology of non-photoionised gas in unprecedented detail at spatial scales where we can disentangle a number of sources (e.g., \hii\ regions, WR stars, shocks etc).

This paper aims to explore chemical homogeneity and understand the ionisation mechanisms responsible for emission within a nearby BCD, Mrk~71 ($D\simeq3.44$~Mpc)\footnote{Here we adopt the Cepheid-derived distance by \citet{Tolstoy:1995}}, and thereby assess the effect of feedback mechanisms on the energetics, structure and star-formation using \textit{HST} narrow-band imaging.  Mrk~71 is an excellent case-study for this, as spectroscopic observations of this galaxy show evidence for both a significant mechanical feedback component and signs of a WR population.  The structure of the paper is as follows:  Mrk~71 is described in detail in Section~\ref{sec:overview}, followed by a description of our observations and data processing in Section~\ref{sec:data}.  We discuss the final images in Section~4, along with information derived from the images, i.e. emission line diagnostics, photoionisation modelling, and metallicity and \heii\ imaging. In Section 5 we  discuss the benefits and limitations of ultra-high spatial resolution observations.  We conclude in Section~6.

\begin{figure*}
\includegraphics[scale=0.7]{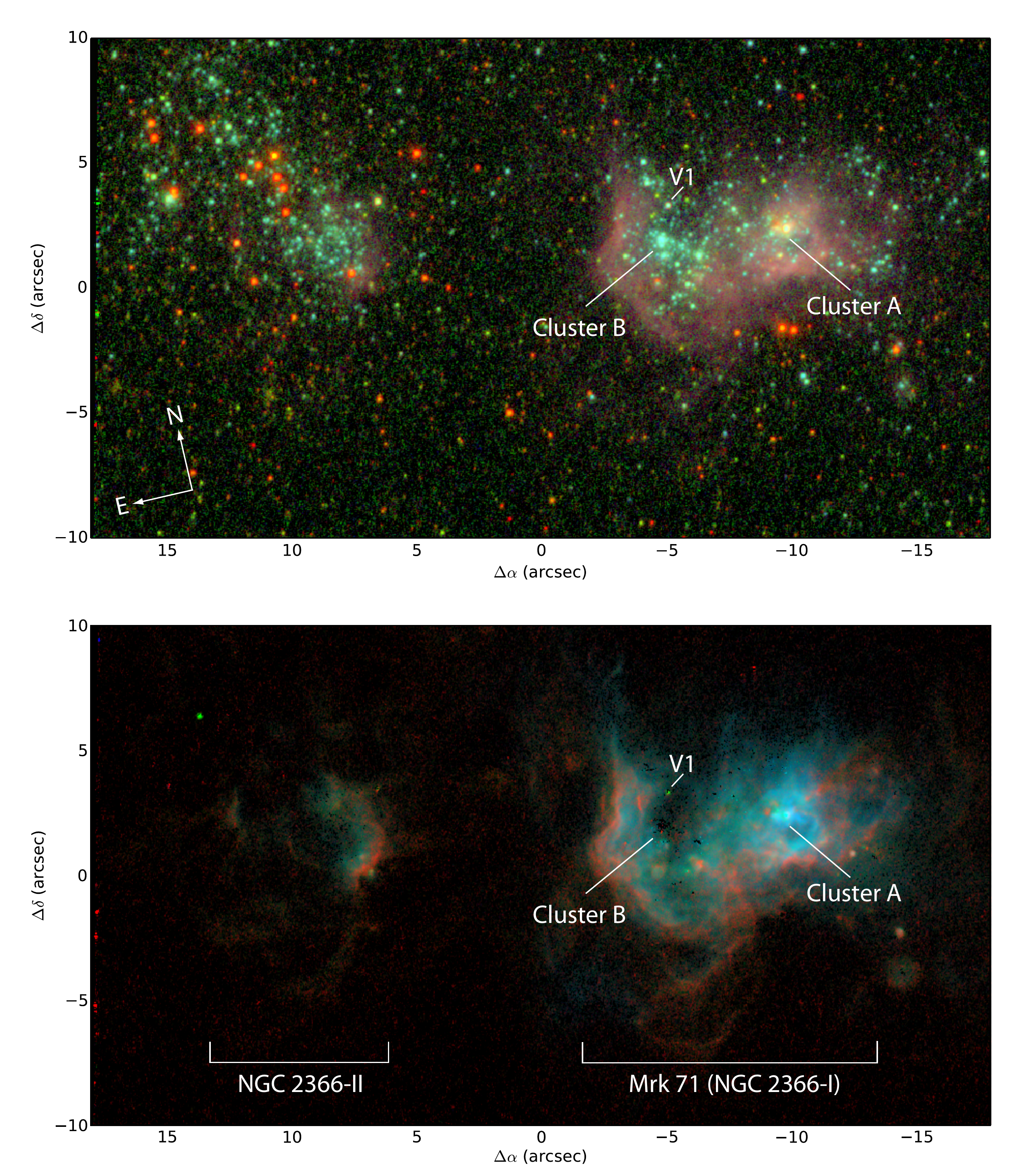}
\caption{Colour composites of Mrk~71 and NGC~2366-II with main clusters (A and B) and LBV star labelled \citep[following the nomenclature of][]{Drissen:2000}.  \textit{Top-panel:}  broad-band images of stellar continuum -  $I$-band (red), $V$-band (green), and $U$-band (blue); \textit{bottom-panel:} narrow-band, continuum-subtracted images of the ionised gas - F373N (\foii, red), F656N (\ha, green), and F502N (\foiii, blue). The images have a spatial scale of $0.04''$/pixel (0.67~pc/pixel) and are $36''\times20''$ in size ($600\times333$~pc).}\label{fig:colour}
\end{figure*}

\section{Mrk~71 - a brief overview}\label{sec:overview}
Mrk~71 is one of the most powerful (local) starbursts known in the (local) Universe and is famed for having the `best case of hypersonic gas' \citep{Roy:1991,Roy:1992,Gonzalez-Delgado:1994}.  Early optical spectroscopic observations by \citet{Kennicutt:1980} revealed extremely high degrees of excitation, low amounts of reddening and an unusually energetic source of ionisation.   Its faint, extended and exceptionally broad ($FWHM>2300$~\kms) emission lines were first reported by \citet{Roy:1992}, thought to be signatures of an expanding supershell.  Owing to this extraordinary feature, this system has been observed from X-ray to far-infrared (FIR) wavelengths, both with imaging and spectroscopy.  Here we provide a brief overview of these observations and studies from the literature.

Structurally, Mrk~71 is a large and massive chain of \hii\ regions within the dwarf galaxy NGC~2366 \citep{Kennicutt:1980,Binette:2009} and is classed as a cometary-BCD.  Following the nomenclature adopted by \citet{Drissen:2000}, the NGC~2366 complex comprises of three regions, I--III, where NGC~2366-I forms Mrk~71 alone.  Within Mrk~71 there are two main clusters, A and B.  Each of these regions are labelled in Figure~\ref{fig:colour}, where we present colour-composite images created from new WFC3 narrow and broad-band images covering both NGC~2366-II and Mrk~71.

The nearby proximity of Mrk~71 ($D=3.44$~Mpc) allows resolved stellar population studies.  Using $HST$, \citet{Thuan:2005} found the stellar population within Mrk~71 to consist of a young ($<30$~Myr) population of blue main sequence stars, an intermediate population of blue and red supergiants (RSGs, 20--100~Myr), an old AGB star population $>100$~Myr, and red giant stars ($>1$~Gyr).  These findings were consistent with \citet{Noeske:2000} who estimated the age of the oldest stars in the low-surface brightness component to be $<3$~Gyr.

With regards to the individual stellar clusters, \citet{Drissen:2000} report that both Mrk~71 and NGC~2366-II show RSGs.  In Mrk~71 itself, cluster A is the younger ($<1$~Myr), more intense starburst with few visible stars, whereas cluster B is slightly older (3--4~Myr) and consists of many dozens of blue stars.  \citet{Drissen:2000} attribute an excess of \heii~$\lambda$4686 emission in cluster B to 5 WR stars, whereas no WR stars are detected in or around cluster A or NGC~2366-II.  Additionally, a rare luminous blue variable (LBV) is also present (labelled `V1'), which has been undergoing an eruption since 1994 \citep{Drissen:1997}.

Chemically, long-slit optical observations show Mrk~71 to be metal-poor, with $\sim$1/6~\Zsol\footnote{Throughout this paper we adopt the solar oxygen abundance of 12+log(O/H)=8.69 from \citet{Asplund:2009}.}  \citep[i.e. 12+log(O/H)=7.89]{Noeske:2000,Thuan:2005,Moustakas:2006}.  The oxygen abundances in several other \hii\ regions along the body of Mrk~71 were found by \citet{Roy:1996} to be slightly higher than the brightest region (cluster A).

The aforementioned broad emission line component of Mrk~71 was observed to extend from cluster A.  Several mechanisms were put forward by \citet{Roy:1992} to explain the emission, including stellar winds, Thompson scattering by hot electrons, supernova remnants, and a super-bubble blow-out (however all were deemed unsatisfactory).  This subject was revisited by \citet{Binette:2009} who demonstrated that the emission could in-fact be due to the interaction of high-velocity cluster winds with dense photoionised clumps.  The super-star cluster responsible for the intense nebular flux observed in cluster A shows no stellar UV absorption features \citep{Drissen:2000}, suggesting that it is shrouded by dust.  Furthermore, the optical--FIR spectral energy distribution of this cluster shows evidence that it is in an ultra compact \hii\ region stage where newly-formed stars are still embedded in their natal molecular clouds.    

The broad emission of Mrk~71 made it an attractive candidate for hosting a low-metallicity AGN \citep{Izotov:2007}. Follow-up X-ray observations by \citet{Thuan:2014} detect four faint point sources coincident with a background AGN and a very the compact \hii\ region in cluster A, and two faint extended sources associated with massive \hii\ complexes.  However, no evidence for hard non-thermal radiation is seen in the optical spectrum of Mrk~71 \citep[][and references therein]{Izotov:2007}.

Fabry-Perot observations by \citet{Roy:1991} revealed emission line splitting across a $240\times 307$~pc region centred on cluster A (i.e. covering the entirety of Mrk~71) which was interpreted as a bubble expanding at a speed of 45~\kms.  Using population synthesis models, \citet{Drissen:2000} show that the expanding bubble must be due to energy released by massive stars and supernovae explosions throughout cluster B, rather than being blown by winds from massive stars within the core of the cluster.  The same radial velocity maps of \citet{Roy:1991} also uncovered a `chimney' structure extending away from the central bubble in the north-northwest direction, which aligns with the obvious cavity in emission seen extending north from cluster B in Figure~\ref{fig:colour} and the diffuse \heii\ extended emission seen by \citet{Drissen:2000}.

Kinematically, \foiii\ and \hi\ velocity maps of NGC~2366 \citep{Roy:1991,Braun:1995} suggest that NGC~2363 ($\sim$1.14~kpc NW of Mrk~71 and not covered here) may be disconnected from the main body of objects (Mrk~71 and NGC~2366-II, i.e. those under study here), and could in-fact be a satellite galaxy which is interacting and perhaps triggering star-formation in the these objects.  This is further supported by the fact that an age-sequence is seen to exist along the main bodies: 10~Myr (NGC~2366-II), 3--5~Myr (Mrk~71-B), $<1$~Myr (Mrk~71-A).  

\begin{table}
\caption{Details of observations used within this study.}
\begin{center}
\begin{tabular}{llc}
\tableline
Filter & Emission line / band & $T_{exp}$/s \\
\tableline
F336W & $U$-band & 300\\
F373N & \foii~$\lambda$3727+$\lambda$3729 & 900\\
F438W & $B$-band & 210\\
F469N & \heii~$\lambda$4686 & 5250\\
F487N & \hb~$\lambda$4861 & 1020\\
F502N & \foiii~$\lambda$5007 & 540\\
F547M & $V$-band & 300\\
F656N & \ha~$\lambda$6564 & 720\\
FQ672N & \fsii~$\lambda$6716 & 2841\\
FQ674N & \fsii~$\lambda$6732 & 2841\\
F814W & $I$-band & 420\\
\tableline
\end{tabular}
\\[1.5mm]
Notes: For each WFC3-UVIS filter we list the corresponding emission line or broad-band emission observed within Mrk~71, along with the total exposure time used.
\end{center}
\label{tab:obs}
\end{table}%

\begin{figure*}
\includegraphics[scale=0.7,angle=90]{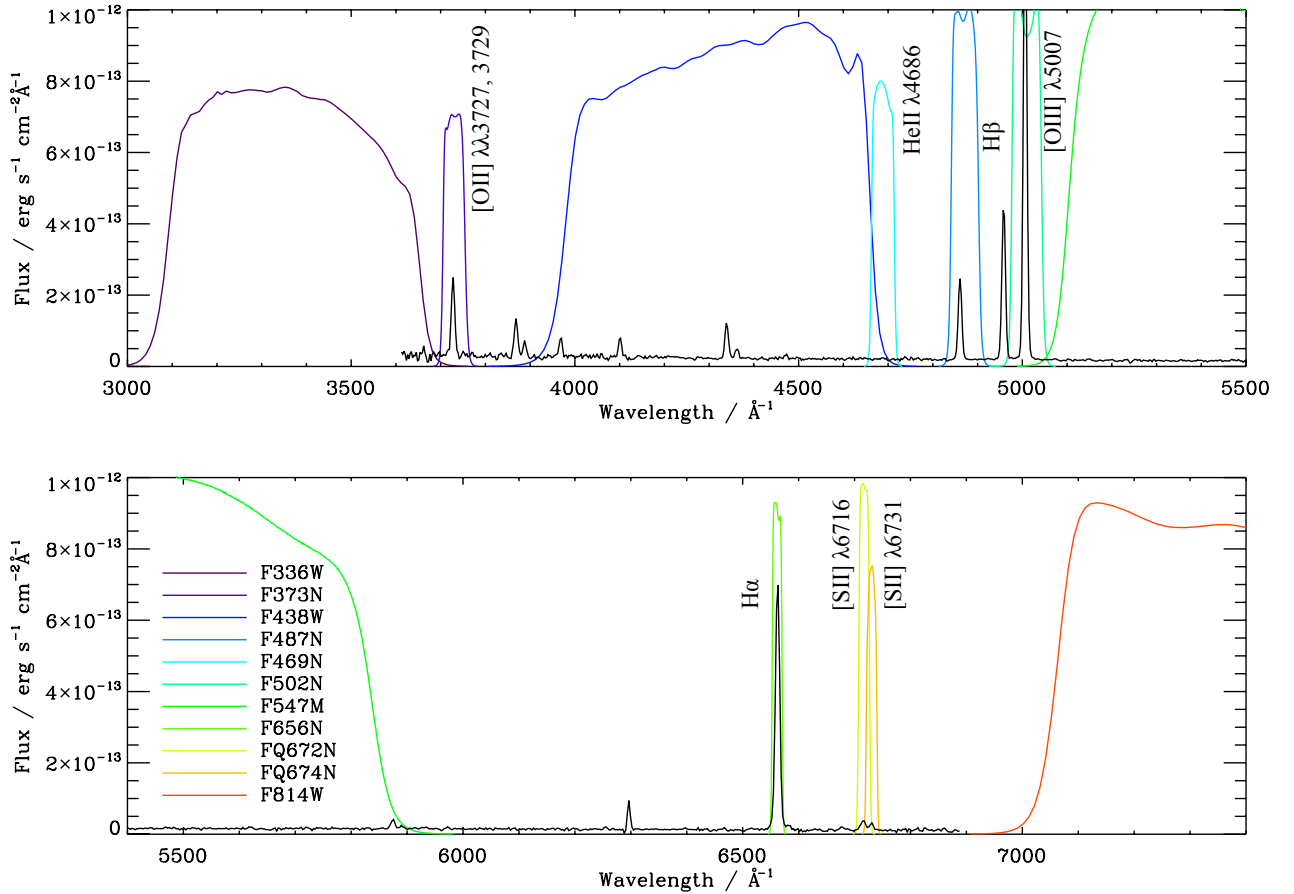}
\caption{Filter transmission curves of each WFC3 filter utilised within this study overlaid on an optical spectrum of Mrk~71 from \citet{Moustakas:2006} (described in Section~\ref{sec:comparison}).  Transmission curves were scaled arbitrarily to match the flux level of the spectrum.}\label{fig:spec}
\end{figure*}

\section{Observations \& Data Processing}\label{sec:data}
The images used in this work are all obtained with the Wide Field Camera 3 (WFC3) UVIS channel on board \textit{HST}, and form part of observational program 13041 (PI: James), observed on 6th March 2013.  A summary of the filters, the emission line or band covered by the filter, and the exposure times is given in Table~\ref{tab:obs}.  Since the focus of this work is on the ionised gas, as probed by strong emission lines, the emphasis will be on the narrow-band filter observations.  The observations in the medium and broad-band filters are primarily used for stellar continuum subtraction, as discussed in Section~\ref{sec:contsub}.  Each narrow-band filter was used to target a specific emission line within the optical spectrum of Mrk~71, which we show in Figure~\ref{fig:spec}.  The optical spectrum\footnote{Available at \url{http://www.sos.siena.edu/$\sim$jmoustakas/\\research/spectral\_atlas/index.html}} was taken from \citet{Moustakas:2006} (MK06 hereafter) and is described below in more detail. 

The field-of-view (FoV) of each WFC3 broad- and narrow-band observation subtends 2.7 kpc$^2$ ($162''\times162''$), whereas quad-band filter observations subtend 0.45~kpc$^2$ (i.e. 1/6$^{th}$ the WFC3 FoV after allowing for edge effects).

The WFC3/UVIS dataset was processed through the \textsc{calwf3} pipeline version 1.0.  The calibrated, flat-fielded individual exposures were corrected for charge transfer efficiency (CTE) losses by using a publicly available stand-alone programme \footnote{Anderson, J., 2013, \url{http://www.stsci.edu/hst/wfc3/tools/cte tools}}.  The resulting files were then aligned and combined using \textsc{drizzlepac} software.  For each filter, individual exposures were aligned using \textsc{tweakreg} and combined using \textsc{astrodrizzle}.  We then aligned the combined images across filters using the F336W image as the reference frame for the WCS.  The final data products have a pixel scale of 4 mas pixel$^{-1}$ and are in units of e$^{-}$\,s$^{-1}$, which were converted to physical units using the WFC3 photometric zeropoints.

\begin{table*}
\caption{A comparison between emission line fluxes measured from WFC3 images and those obtained from spectroscopic studies within the literature (described in Section~\ref{sec:comparison}).}
\begin{center}
\begin{tabular}{l c cc@{\hspace{1cm}} ccc }
\tableline
 &	WFC3 Images - total	&  \multicolumn{2}{c}{{\hspace{-1cm}}MK06}		&	\multicolumn{3}{c}{TI05}						\\
\cline{2-2} \cline{3-4} \cline{5-7}
Line ID &	F$_\lambda$	&	F$_\lambda$	&	\%\ agreement	&	F$_\lambda$		&	F$_\lambda$ images	&	\%\ agreement	\\
\tableline														\\
\foii&	129.00	&	142	&	91	&	25.12		&	22	&	88	\\
\heii &	4.59	&	--	&	--	&	0.45		&	0.485	&	108	\\
\hb &	176.00	&	220	&	80	&	51.95		&	46.6	&	90	\\
\foiii &	1,130.00	&	1,270	&	89	&	355.16	&	326	&	92	\\
\ha &	604.00	&	650	&	93	&	--			&	--	&	--	\\
\fsii &	18.7	&	43.8	&	43	&	--			&	--	&	--	\\
\tableline
\end{tabular}
\\[1.5mm]
Notes: The spectroscopic results of  \citet{Moustakas:2006} were obtained by integrating over the entire galaxy and are therefore compared against fluxes summed over the entire WFC3 image.  The long-slit observations of \citet{Thuan:2005} (TI05) were centred on the brightest region in Mrk~71.  In this case we list the fluxes measured on the images within a simulated aperture (F$_\lambda$ images). Observed fluxes (F$_\lambda$) are in units of $1\times10^{-14}$ erg/s/cm$^2$.
\end{center}
\label{tab:fluxes}
\end{table*}%

\subsection{Continuum Subtraction}\label{sec:contsub}

In order to create emission-line-only images, narrow-band images were
corrected for underlying stellar continua.  The nebular emission lines
contributing to the flux in each of the seven narrow-band filters are listed in
Table~\ref{tab:obs}.  Each narrowband filter targets a
single major emission line (as illustrated in Figure~\ref{fig:spec}).  In
particular, F656N exclusively covers \ha\, and only negligible contribution from \fnii~$\lambda$6584 is
expected.   Nebular lines do not provide any significant flux in the F336W and F547M filters, but there is a small contribution to the F814W filter (predominantly from \fsiii\ and the Paschen series). To determine what effect this might have on the continuum subtraction we use the near-infrared spectrum presented in \citet{Izotov:2011b} to quantify the nebular line contribution to the F814W flux. We find that less than 20\%\ of the flux is due to nebular emission, and using a scaling based on the \ha\ image and repeated continuum subtraction, the \fsii\ flux is increased by less than 0.02~dex.

Traditionally, narrow-band images are corrected for the stellar continuum using
linear interpolation and extrapolation from neighbouring continuum (i.e.
broad-band) images.  However, here we adopt an alternative approach that exploits the wide wavelength coverage of the set of broad-band filters to-hand. We use the F336W, F547M, and F814W images to determine a spatially-resolved stellar SED by fitting a linear combination of \citet{Bruzual:2003} stellar population models to each pixel. We then calculate and subtract the contribution of these SEDs to the emission in each of the narrow-band images, leaving just the (presumed) nebular emission. Finally, to determine the emission line fluxes we multiply normalised Gaussian SEDs by the narrow-band filter transmission functions and divide these `normalised line fluxes' into the continuum-subtracted images to calculate the actual amplitudes of the emission lines. We have done this assuming a range of line widths from 70~\kms\ to 150~\kms\ and find that the fluxes vary by only a few per cent. To investigate the robustness of our continuum subtraction we have also performed the continuum fitting using \textsc{STARBURST99} models \citep{Leitherer:1999}, which include nebular continuum emission, and find negligible changes to the calculated emission line fluxes.

A colour-composite image from the broad-band filters is shown in Figure~\ref{fig:colour} alongside an image made from three emission line images, which illustrates that our method has done an excellent job of removing the stellar emission.\footnote{It should also be noted that continuum subtraction was also performed using the method of \citet{Hong:2014} and residual structure surrounding stellar emission in the continuum-subtracted images was seen to be larger than those created with the SED-method described here.} We note that this methodology does not account for differences in the point spread function between different filters, but this will not affect our subsequent analysis as we re-bin the data to be larger than the point spread function. 

Hereafter each continuum-subtracted narrow-band image will be referred to in terms of the emission line flux that they cover, i.e. \foii, \heii, \hb, \foiii, \ha\ and \fsii\, with the latter being the sum of the two \fsii\ filters.

\subsubsection{Accounting for emission line structure within the filter}
As discussed in Section~\ref{sec:overview}, Mrk~71 is known to harbour both very broad emission lines and kinematical structure, two properties which can affect whether the emission lines are being fully covered by each of the filter band-passes.  In order to asses the magnitude of these effects on our measured emission line fluxes,  we simulated emission lines for a range of velocities and widths that represent the `worst-case-scenario', and measured the change in total flux seen within each narrow-band filter.  Velocities were chosen to range from 0--160~\kms, which accounts for both the systemic velocity of 90--100~\kms\, measured from optical emission line spectra \citep[e.g. MK06 and ][]{Roy:1991} and the \hi\ velocity maps \citep{Hunter:2012}, and an expansion velocity of 45~\kms\ measured by \citet{Roy:1991} for the bubble surrounding cluster A.  The model emission lines consisted of a narrow component, with $FWHM=110$~\kms\ and a broad component with $FWHM=2400$~\kms, as measured by \citep{Roy:1992} and \citet{Izotov:2007}, with the broad-component being observed only in the \foiii~$\lambda\lambda4959,5007$ and Balmer lines. Changes in the velocity had a negligible effect on the emission lines in all cases, i.e. $<1$\%.  The emission line width had a larger effect, most notably for \ha\ because the F656N filter is particularly narrow in order to miss \fnii~$\lambda$6584.  However, since the broad-component emission is known to contain $<2$\%\ of the total emission line flux for each line \citep{Izotov:2007}, this effect is also negligible on our measured emission line fluxes.

\begin{figure*}
\includegraphics[scale=0.47]{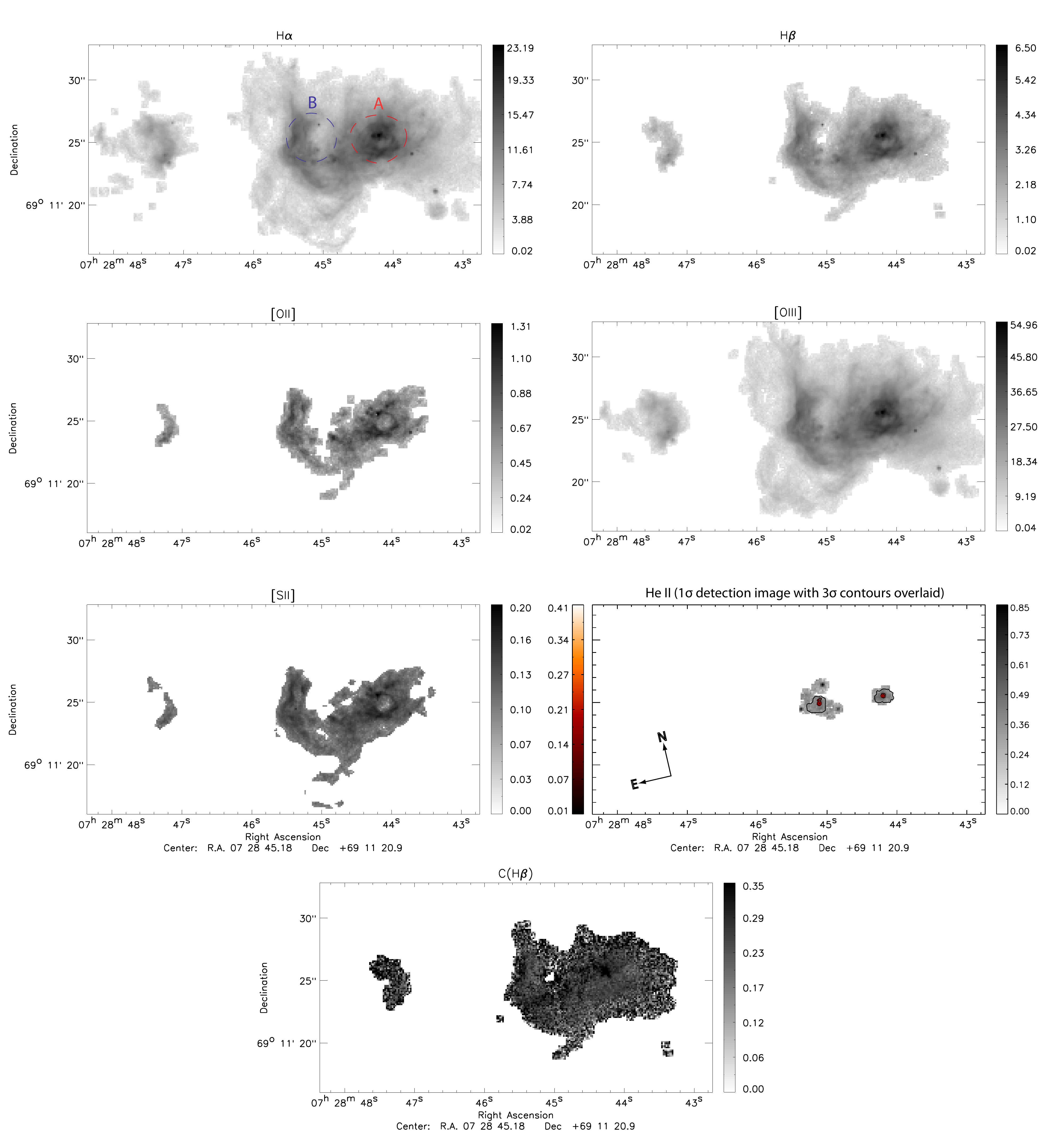}
\caption{Continuum-subtracted emission line images of Mrk~71 corresponding to the narrow-band observations listed in Table~\ref{tab:obs}.  Fluxes are in units of $1\times10^{-15}$ erg/s/cm$^2$.  Each map shows 3$\sigma$ pixel detections after 3$\times$3 pixel binning, with the exception of \heii, which shows 1$\sigma$ detections after 3$\times$3 pixel binning.  We overlay the 3$\sigma$ detected \heii\ flux as contours in the \heii\ panel (with 50\%\ of the peak flux masked).  The bottom centre panel shows the C(\hb) image of Mrk~71, used to correct each narrow-band image for reddening when necessary.  Each emission line image is normalised to its peak flux. Apertures overlaid on the \ha\ image denote the spatial extent of regions A and B which are referred to in Section~\ref{sec:results}.}\label{fig:nbands} 
\end{figure*}

\subsection{Comparison of fluxes with literature}\label{sec:comparison}
In order to test the accuracy of our continuum-subtraction method, the fluxes in the final, emission-line images were compared with fluxes from ground-based spectra from the literature.  Two comparison spectra are available to us: (i) integrated spectrophotometry from MK06, where fluxes are obtained via a drift-scanning technique and integrated over 90$\times$60~arcsec$^{2}$ galaxy (Mrk~71=ID 96 within their study).  These fluxes provide the most accurate comparison with the total flux integrated over each image, i.e. removing any possible slit-loss effects.; (ii) MMT optical long-slit observations (extracted over $6''\times2''$ aperture centred on the brightest part of the galaxy) published by \citet{Thuan:2005} (TI05 hereafter).  Here the authors, extract fluxes from three different \hii\ regions in the same galaxy, although the specific positions of these regions and the position angle (PA) of the slit are not given.  As such, we use the summed fluxes from all three regions for comparison with a simulated $6''\times2''$ aperture centred on the brightest region of Mrk~71. 

The fluxes from each of the observations of MK06 and TI05 are shown in Table~\ref{tab:fluxes}.  For each observation, we also list the flux measured from our emission-line images using the same extraction aperture (to our best knowledge).  It can be seen that we are in very close agreement with the spectroscopic fluxes from the literature, obtaining $>80\%$ in all emission-line images, with the exception of \fsii\ for which there appears to be a discrepancy by a factor of two.  As the two observations represent both the total flux (MK06) and the core flux (TI05) within Mrk~71, and a good agreement is found in both cases, we are confident that both our flux calibration and continuum subtraction have been successful in creating reliable emission-line images.

\section{Results} \label{sec:results}
Before beginning our analysis, emission-line images were binned by $3\times3$ pixels.  The final pixel scale is 0.12~\arcsec, or $\sim$2~pc.  Using the uncertainty in each pixel, we clip each image by 3$\sigma$ in order to keep the highest S/N pixels (and most reliable information), with the exception of \heii\ where we clip by 1$\sigma$ (this line is particularly faint and any deeper rejection cuts would remove most of the `WR' signature emission).  

The final, continuum-subtracted images in emission lines \ha, \hb, \foii, \foiii, \heii, and \fsii\ are shown in Figure~\ref{fig:nbands}.  Both NGC~2366-II and Mrk~71 are seen in all emission line images, with the exception of \heii\ which is only observed in Mrk~71.  Cluster A (a super star cluster) is clearly defined in the western part of Mrk~71 with a shell of surrounding emission.  The ionised gas within several smaller clusters are also seen throughout Mrk~71, as traced by the majority of the emission lines.   The eastern part of Mrk~71, containing cluster B, is $\sim9''$ ($\sim150$~pc) away from cluster A and has a sharp edge of emission $\sim0.5''$ ($\sim8$~pc) wide along its eastern side, suggestive of a shock-front.  A large `arm' of emission is seen to extend north from cluster B, $\sim2''$ ($\sim45$~pc) in width.   The emission in \ha, \hb, and \foiii\ is seen to extend further than that of \foii\ and \fsii.  We discuss \heii\ emission in detail in Section~\ref{sec:heii}.  

Also shown in Figure~\ref{fig:nbands} is the C(\hb) image, i.e. extinction image created from the \ha/\hb\ Balmer line ratio and adopting the case B recombination ratio F(\ha)/F(\hb)=2.87 (appropriate for gas with \elt=10,000~K and \eld=100~\cm3), and the Large Magellanic Cloud extinction curve \citep{Fitzpatrick:1999}.  It should be noted that any stellar photospheric absorption affecting the Balmer lines was corrected for during the SED-based continuum subtraction (\S~\ref{sec:contsub}).    On average, we find the Balmer decrement implies an average C(\hb)$=0.13\pm0.04$ ($E(B-V)=0.18\pm0.06$).  In order to minimise the introduction of noise within the emission-line images, we only correct for reddening when necessary (i.e. when ratioing emission line images that differ substantially in wavelength).  

\begin{figure*}
\includegraphics[scale=0.47]{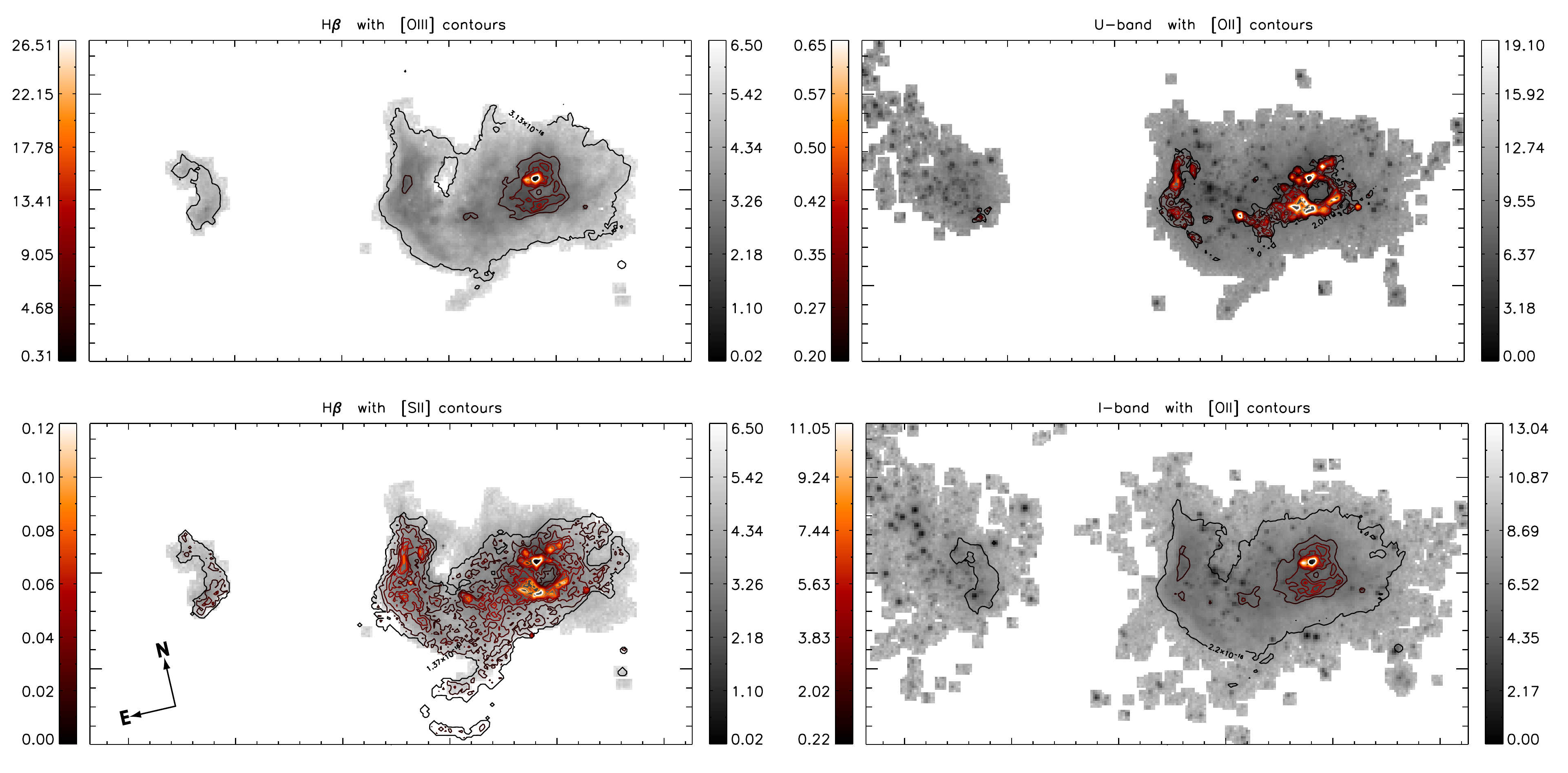}
\caption{\textit{Left-panels:} A comparison between the morphology of Balmer line and forbidden line emission throughout Mrk~71: \hb\, emission line image with \foiii\ \textit{(top)} and \fsii\, \textit{(bottom)} contours overlaid. \textit{Right-panels:} The emission line morphology of Mrk~71 in relation to the stellar continuum emission:  $U$-band image with \foii\ contours overlaid \textit{(top)}; $I$-band image with \ha\ contours overlaid  \textit{(bottom)}.  For each contoured image we mask 50\%\ of the peak flux in order to show the low-level forbidden-line emission.}\label{fig:conts}
\end{figure*}

In Figure~\ref{fig:conts}  we compare the morphology of emission in the forbidden lines, \foiii\ and \fsii\, with that of the Balmer line emission, \hb.  In each case it can be seen that the overall morphology of the two emission lines closely traces one another, with peaks in flux surrounding cluster A and along the eastern edge of the arm, although the extent of  \hb\ and \foiii\ emission is larger than \fsii.  The morphology of \foiii\ and \fsii\ differ somewhat in the regions surrounding cluster A, where the lower ionisation line, \fsii, shows a pronounced shell around the cluster and the \foiii\ emission is more localised.  This could be interpreted as seeing the edges of an ionised bubble of gas, $\sim$22 pc in diameter,  which can be thought of as a single \hii\ region or Str{\"o}mgren sphere.  However, the main ionising source in cluster A is not seen in the centre of this `bubble' - it instead lies on the northern edge of the ring (i.e. coincident with the strongest peak in \foiii).  This scenario is in line with \citet{Drissen:2000}, who perceive cluster A to be a dense star cluster still embedded in dust and in its early stages of formation (i.e. $<1$~Myr).

Cluster A's location within the periphery of the bubble, combined with its young age and compact nature, may imply that this is a localised region of star-formation that was triggered by bubble itself, i.e. representing a second generation of star-formation. Although there are no signs of an older cluster of star-formation within close proximity to the bubble (Figure~1), its existence cannot be ruled out as it may have simply dispersed. However, we can address this scenario via the age of the bubble.  Following the methodology of \citep{Martin:1998}, if we assume a typical expansion velocity of 50~\kms, the bubble would have a dynamical age of $\sim$0.3~Myr, i.e. quite in line with the age of cluster A.  Also, in order for the bubble to have triggered a 1~Myr old cluster, it would need to be traveling at $<<10$~\kms. Overall, in consideration of the short time scales involved in both cluster A and the bubble, the most probably scenario is that they are associated with one another.

\begin{figure*}
\includegraphics[scale=0.5]{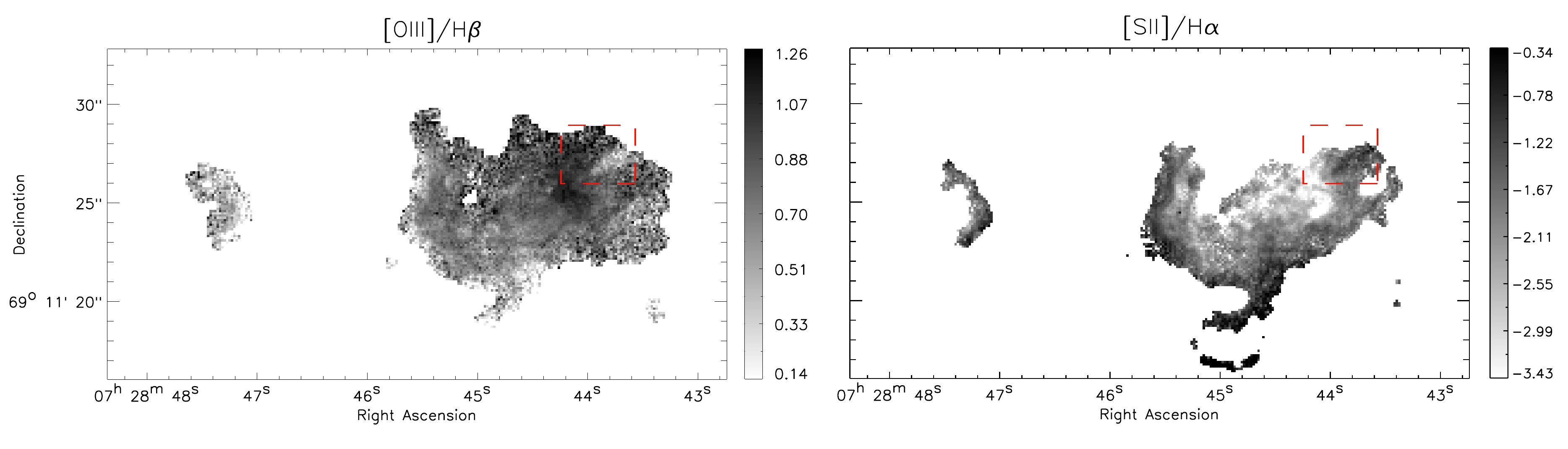}
\caption{Imaging BPT emission line diagnostics across Mrk~71: the emission line ratio of \foiii/\hb\ \textit{(left)} and \fsii/\ha\ \textit{(right)}. Red dashed box highlights a region of highly ionised gas alongside low-ionised gas, i.e. structure suggestive of an ionisation front (see text for details).}\label{fig:BPT_im}
\end{figure*}

\begin{figure*}
\includegraphics[scale=0.4]{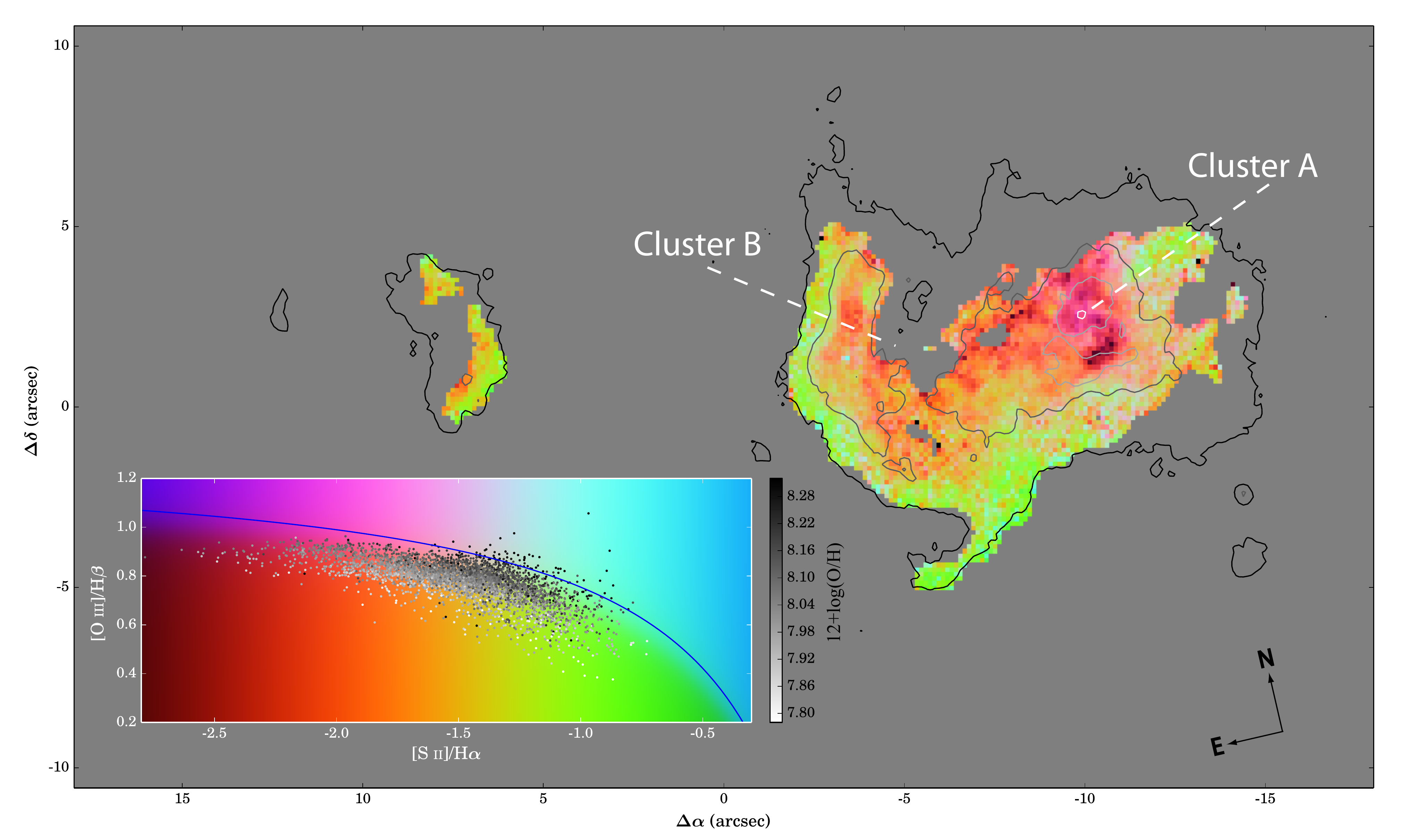}
\caption{An emission line diagnostic map showing log(\fsii/\ha) vs. log(\foiii/\hb), derived from the line ratio images in Fig.~\ref{fig:BPT_im} and colour-coded according to their location within the BPT diagram shown in the inner panel.  Overlaid contours represent the morphology of \ha\ emission (Fig.~\ref{fig:nbands}).   The grey-scale of the data points overlaid in the BPT diagram (inner panel) indicates their metallicity (as discussed in Section~\ref{sec:met}).  The `maximum starburst line' from \citet{Kewley:2001} is overlaid in blue. }\label{fig:BPT_col}
\end{figure*}

\begin{figure*}
\includegraphics[scale=0.7,angle=90]{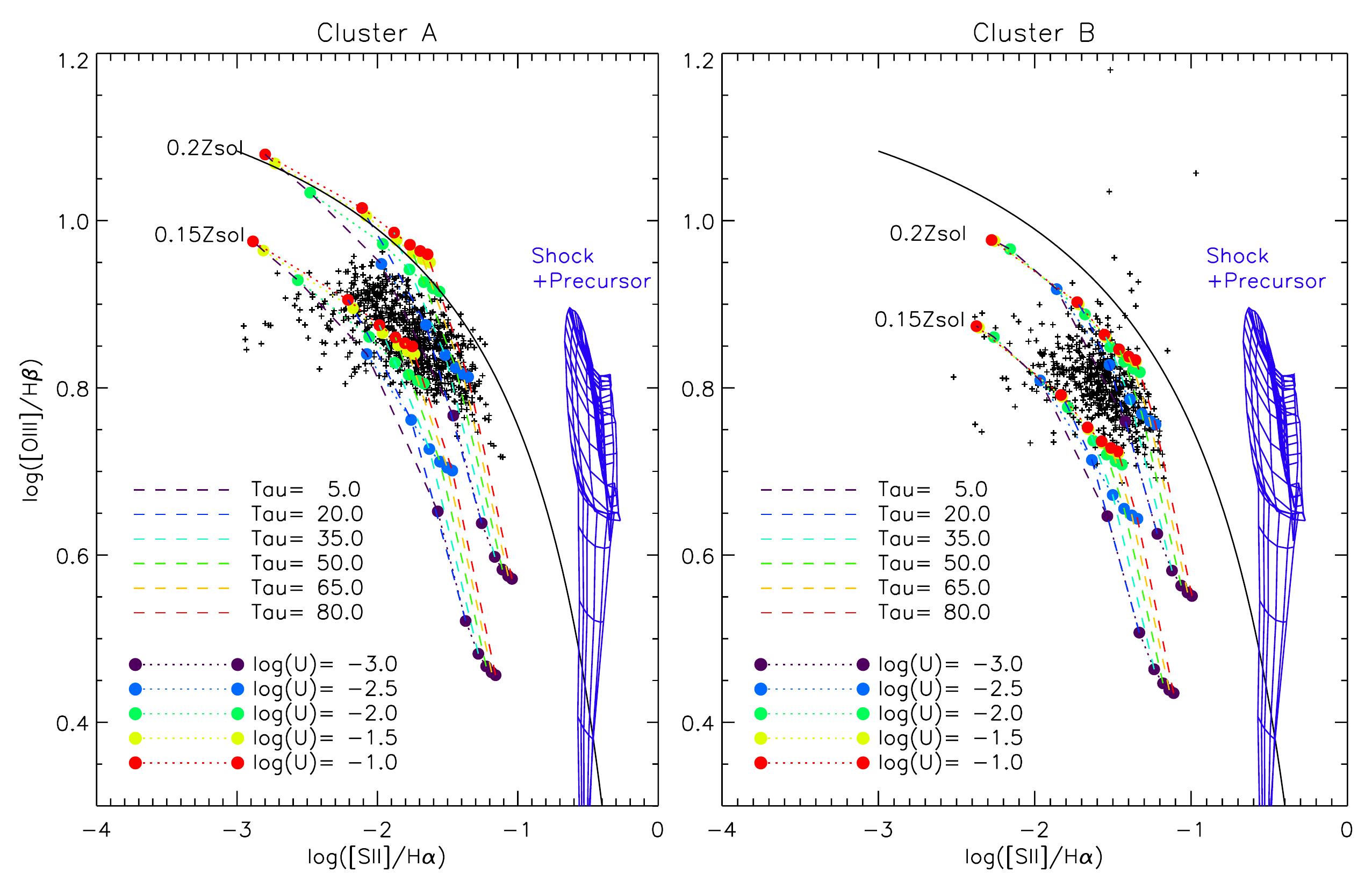}
\caption{ Pixels corresponding to the separate regions of gas surrounding clusters A (left panel) and B (right panel).  The spatial extent of these regions are denoted in Fig.~\ref{fig:nbands}.  Overlaid are the customised photoionisation model grids described in Section~\ref{sec:models}, which cover 0.15~\Zsol\ and 0.20~\Zsol, ionisation parameters of $log(U)=-3$ to $-1$ and optical depths of $\tau=3$ to 103.  We also overlay the shock+precursor models at LMC metallicity from \citet{Allen:2008} (blue solid lines) and the `maximum starburst line' from \citet{Kewley:2001} (black solid line).}\label{fig:mod}
\end{figure*}

\subsection{Emission Line Diagnostic Imaging}\label{sec:BPT}
The line diagnostic diagram from nebular line emission, \foiii~$\lambda$5007/\hb\ versus \fsii($\lambda6716+\lambda6731$) (or \fnii~$\lambda$6584/\ha), has been used to study the starburst or active galactic nucleus activity in galaxies \citep[][K01 hereafter]{Baldwin:1981,Kewley:2001} and to investigate the ionised gas structure of the ISM for resolved regions \citep[e.g.][]{Calzetti:2004,Westmoquette:2007,Hong:2011}.  These emission line diagnostics are successful in separating different sources of ionisation because the relative intensity of the two collisionally excited lines, \fsii\ and \foiii, depends on the shape and strength of the ionising radiation (along with the metallicity of the ISM), whereas the Balmer recombination lines have the same dependence on the photoionisation process and are used only to normalise out the ionising luminosity.  For example, in photoionised gas, as the number of high energy photons increases, there is an increase in \foiii\ emission and a decrease in \fsii\ emission (due to the increase in doubly ionised sulphur).  This effect can be seen in Figure~\ref{fig:BPT_im}, where we show Mrk~71 in pixel-by-pixel \fsii/\ha\ and \foiii/\hb\ emission line ratios.  In the gas surrounding cluster A there is a peak in \foiii/\hb, whereas \fsii/\ha\ can no longer be seen.  Then moving away from the strong photoionising emission from the OB stars within cluster A,  \fsii/\ha\ increases while \foiii/\hb\ decreases.  

Interesting additional structure can also be seen in these emission line ratio images.  Firstly, directly above cluster A, a strip of highly-ionised gas can be seen directly alongside a strip of low-ionised gas (i.e. decreased \foiii/\hb\, as highlighted with red-dashed box), which is also mirrored in the \fsii/\ha\ image.  This structure is suggestive of an ionisation front from highly-ionised gas being blown away from the hot OB stars in the main cluster.  Secondly, the eastern edge of the arm is traced by an increase in \fsii/\ha\ and a decrease in \foiii/\hb, signifying a sharp transition between high--medium ionised gas.  A similar structure can also be seen along the edge of NGC~2366-II.  The morphology and sharp gradients of these emission-line ratios suggest the existence of shocked gas along the eastern edge of the arm as the gas moves away from cluster B and shocks against the surrounding ISM. \looseness=-2

In order to explore the excitation mechanisms fully, in Figure~\ref{fig:BPT_col} we plot both line ratio images in a colour-coded diagnostic parameter space as \fsii/\ha\ vs. \foiii/\hb. Also shown is the `maximum starburst line' from K01, above which the line ratios cannot be due to photoionisation alone, demonstrating that the large majority of the gas within Mrk~71 and NGC~2366-II is in-fact photoionised.  Alternatively, the photoionised emission may be dominating over shock-excitation, despite the 2~pc/pixel resolution.  However, if the super-bubble surrounding Mrk~71 is expanding at only 45~\kms\ \citep{Roy:1991}, the lack of shock-excited emission is perhaps not surprising. We additionally colour-code the pixels overlaid in the BPT diagram according to their metallicity (discussed in Section~\ref{sec:met}).  A gradient in metallicity can be seen in the direction of \foiii/\hb\ and follows the shape of the maximum-starburst line.   Figure~\ref{fig:BPT_col} contains a wealth of information about the ionisation throughout Mrk~71. Firstly, we can clearly see that Mrk~71 has a very inhomogeneous ionisation structure, where cluster A occupies a region of higher \foiii/\hb\ and lower \fsii/\ha\ than cluster B, and the lowest \foiii/\hb\ and \fsii/\ha\ lie in the galaxy outskirts.  
The morphology of the strong \foiii/\hb\ line ratio in cluster A aligns with the peak in \ha\ emission (overlaid grey-scale contours).  
The gradient of decreasing \foiii/\hb\ as you move away from cluster A appears to support the findings of \citet{Drissen:2000}, who suggest that cluster A provides the bulk of the ionising flux for the entire of Mrk~71.  However, while the gradient is strong in the west-to-south direction from cluster A, it is significantly more shallow in the opposite direction, i.e. towards cluster B.  This suggests that the gas throughout Mrk~71 may be receiving a non-negligible contribution of ionising photons from Cluster~B also. The ionising source and physical conditions of both clusters is explored further in the following section. \looseness=-2

\subsubsection{Photoionisation models}\label{sec:models}
We investigate the properties of the ionised gas further using \textsc{mappings v}\footnote{Available at \url{miocene.anu.edu.au/Mappings}} (Sutherland et al., in-prep).  Separate models for clusters A and B are constructed in order to account for the different stellar populations within them (Section~\ref{sec:overview}) and subsequent chemical and physical conditions.  A number of constraints were used to tailor the model for each cluster using both the data presented here and information from previous studies.  
  
For each cluster, we run a grid of spherical, isobaric \hii\ region models at metallicities of 0.15~\Zsol\ and 0.2~\Zsol\ \citep[which represent the elemental abundances measured by][where abundances are given for both clusters separately]{Izotov:1997}.  We adopt ionising luminosities of $\log(L_{H\beta})=39.2$ and $38.7$~ergs\,s$^{-1}$ for clusters A and B, respectively (as measured from our reddening-corrected \hb\ images using the cluster apertures shown in Figure~\ref{fig:nbands}).  In order to match the \heii/\hb\ ratios of each cluster ($(9.34\pm6.06) \times10^{-3}$ and $ (7.81\pm5.16)\times10^{-2}$ for clusters A and B, respectively) it was necessary to use input spectra for non-LTE stellar atmospheres from \citet{Rauch:2003} with effective temperatures of 90 and 105~kK, respectively.  
 
Grids were initially run for pressures between $P/k=10^5$ to $10^8$, where $k$ is the Boltzmann constant, until the electron density of each cluster was matched.  The range in pressure is equivalent to that of density, as the temperature remains fixed at $T=1\times10^4$~K. The electron density (\eld) spatial profiles from \citet{Perez:2001} using \fsii~$\lambda6716/6731$ show averages of \eld\ $=300$~cm$^{-3}$ and 100~cm$^{-3}$ for clusters A and B, respectively, which were reproduced by models at $\log(P/K)=7$ and 6.5, respectively.
 
Once the pressures were constrained for each cluster,  the inhomogeneous nature of the nebular gas seen in Figure~\ref{fig:BPT_col} was recreated by running each metallicity model for a range of the ionisation parameters, $U$ (which represents the density ratio of ionising photons to particles), between $\log(U)=-3$ to $-1$ in steps of $\Delta U=0.5$ and terminating them between a range of optical depths ($\tau$) between $\tau=5$ to 80 in steps of $\Delta\tau=15$.  The range of ionisation parameters was chosen to encompass the average $U$-parameter for each cluster, $\log(U)=-1.89\pm0.3$ and $-2.12\pm0.23$ for clusters A and B, respectively, derived from the \foiii/\foii\ ratio image (discussed and shown in Section~\ref{sec:heii}) and the metallicity of each cluster using the relations of \citet{Kewley:2002}.   

We overlay the photoionisation grids on BPT diagrams in Figure~\ref{fig:mod}, which show both clusters on a pixel-by-pixel basis.  For comparison with non-photoionisation excitation, we also show the emission line ratios predicted by the shock+precursor models generated by \textsc{mappings III} by \citet{Allen:2008}, at LMC metallicity (the closest in metallicity to that of Mrk~71).  A minimum offset of $\sim$0.7~dex is seen between the observed \fsii/\ha\ emission line ratios and the shock models.  As this cannot be accounted for by the possible over-subtracted continuum in the \fsii\ image (as discussed in Section~\ref{sec:contsub}), it appears that the gas within Mrk~71 is predominantly photoionised.  For both clusters, a large majority of line ratios are successfully reproduced by the $U$ and $\tau$ grids between 0.15 and 0.2~\Zsol, i.e. the gas appears to be dense, with $\log(U)=-2.5$ to $-1$, and ranging from optically thin to thick.  In cluster A, the points with $\log$(\fsii/\ha)$<-2$, which are clustered in a shell-like structure (Fig.~\ref{fig:BPT_col}), lie at the optically thin and low-metallicity end of the grid, suggesting that this region should have a particularly high surface brightness.  The points that lie below the grids could be due to the fact that we are integrating through a line-of-sight off the centre of the cluster rather than directly out (like the photoionisation models). In this scenario, it is possible to see an \foiii/\hb\ ratio which is characterised by a larger effective optical depth, while still missing the lower ionisation region entirely and showing lower \fsii/\ha\ ratios.  For cluster B the model grid lies at lower \foiii/\hb\ line ratios as it is generated at a slightly lower pressure (= electron density) and the points do not extend down to the low \fsii/\ha\ values observed in cluster A.  Our best-fit models suggest that both clusters are dense regions of intense star-formation, with metal content ranging from 0.15--0.2~\Zsol\ and a clumpy, dusty medium.  These findings are in agreement with the UV models of \citet{Drissen:2000} that suggest cluster B is 2.5--5 ~Myr old and contains a population of massive O and B-type stars, while cluster A is younger ($<1$~Myr), in the ultracompact \hii\ region stage and shrouded by dust, where newly formed stars are still embedded in their natal molecular clouds. 
\newpage
\subsection{Metallicity Imaging}\label{sec:met}
\begin{figure}
\includegraphics[scale=0.4]{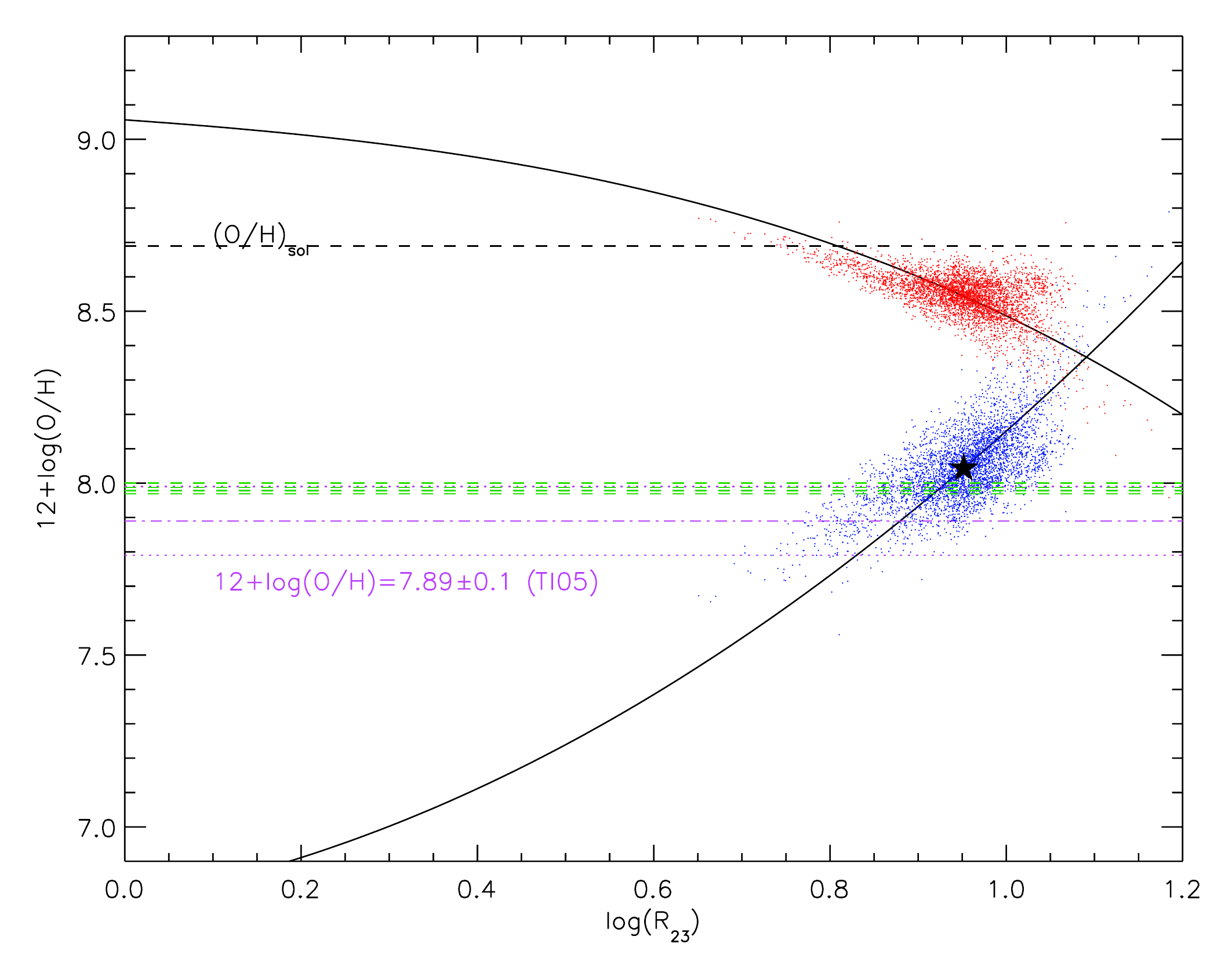}
\caption{The $R_{23}$ emission line ratio as a double-branched function of metallicity (solid-black lines).  The overlaid red and blue points are pixels in the R$_{23}$ (i.e. (\foiii + \foii)/\hb) emission-line image of Mrk~71, corresponding to `upper-' and `lower-branch' metallicities, respectively.  Also shown is the direct-method metallicity and its uncertainty derived from long-slit spectroscopic observation of TI05 (purple dot-dashed and dotted lines, respectively), which suggests that the metallicity distribution of Mrk~71 lies on the lower branch.  The mean metallicity derived from our lower-branch $R_{23}$ images is represented by a black star.  Green dashed lines indicate the $R_{23}$ lower-branch metallicities derived from the emission-line fluxes of TI05 (long-slit) and MK06 (integrated).}\label{fig:R23}
\end{figure}

Our extensive set of narrow-band imaging also allows us to investigate the distribution of metallicity (i.e. oxygen abundance, O/H) throughout Mrk~71 and NGC~2366-II.  For this we use the $R_{23}$ index, an index first proposed by \citet{Pagel:1979} and now the most widely-used approximate indicator of the oxygen abundance in star-forming galaxies, based on the ratios of what are typically the strongest emission lines from \hii\ regions at visible wavelengths: $R_{23} \equiv [I(3727) + I(3729) + I(4960) + I(5008)]/I(H\beta)$.  

One drawback of this index is that a reddening correction is required, since the lines in question are spread over $\sim$1300~\AA.  We therefore correct each emission line image using the extinction image shown in the bottom centre panel of Figure~\ref{fig:nbands}. In deducing the metallicity distribution from the de-reddened emission-line images, we made use of the analytical expressions by \citet{McGaugh:1991} as given by \citet{Kobulnicky:1999}.  These expressions take into account the effect of the ionisation parameter on the relationship between $R_{23}$ and O/H by including a term which depends on the ratio of the \foii\ and \foiii\ lines, $O_{32}\equiv[I(4960) + I(5008)]/[I(3727) + I(3729)]$. 

\begin{figure*}
\includegraphics[scale=0.7]{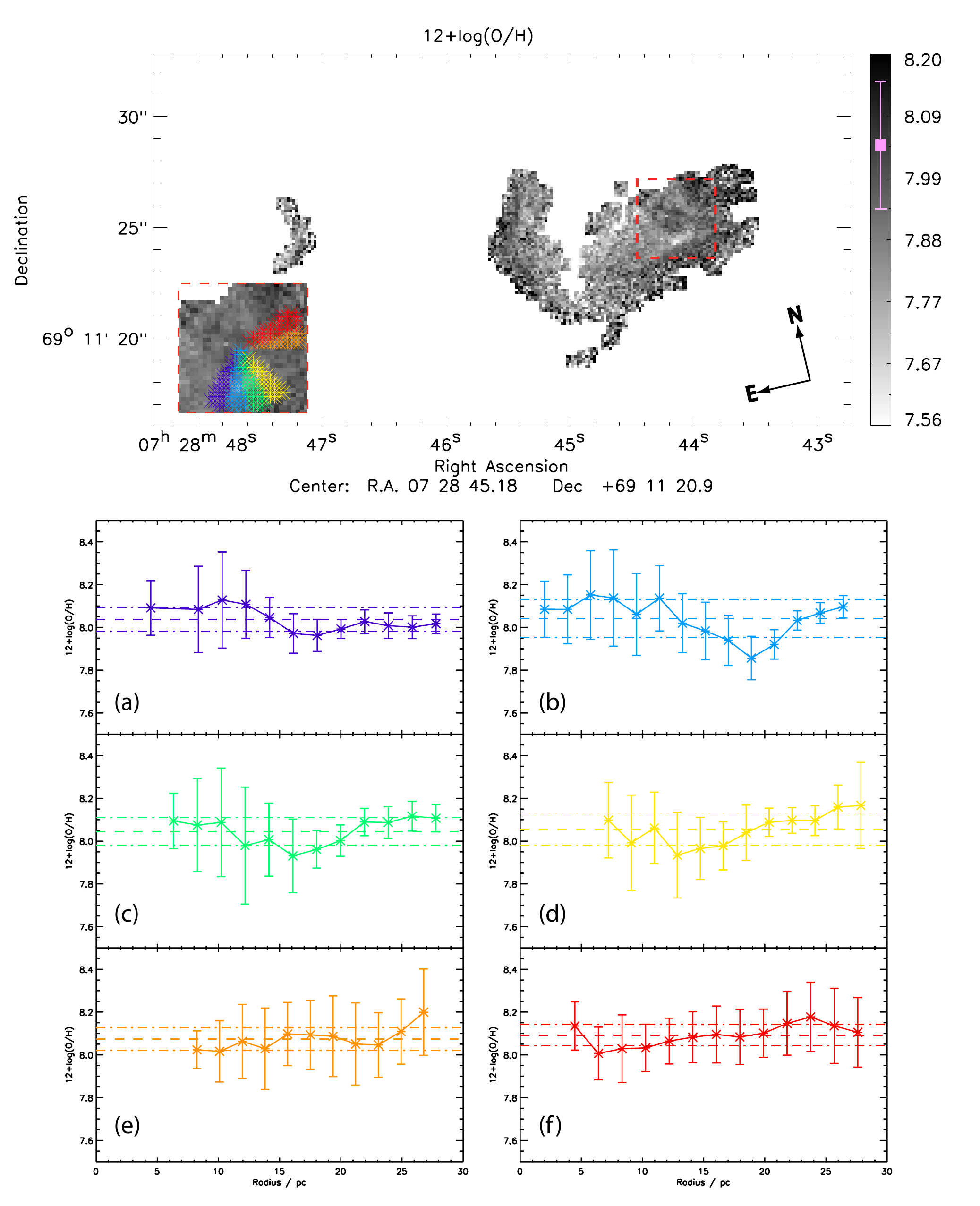}
\caption{\textit{Top-panel: }A `metallicity image' of Mrk~71, corresponding to the lower-branch of the $R_{23}$ diagnostic.  The pink data point shown within the colour bar represents the average metallicity and 1$\sigma$ distribution derived from the image, $<12+\log$(O/H)$>=8.04\pm0.11$. \textit{Lower-panels: } Radial cuts from the centre of the highlighted region (red-dashed box and also shown on a larger scale in inner panel).  Each profile represents the average metallicity in each radial bin along each coloured segment shown in the inner panel, and incorporates the total uncertainty in 12+log(O/H).  Each segment's average metallicity and 1$\sigma$ distribution are shown overlaid as dashed and dot-dashed lines, respectively. }\label{fig:met}
\end{figure*}

Another, and perhaps the most major, drawback of this index is that the relationship between $R_{23}$ and O/H is double valued, so that other line ratios are required to break the degeneracy.   Unfortunately, other metallicity-dependent line ratios such as \fnii/\ha\ \citep{Pettini:2004}, are not available here.  However, we do have O/H measurements from the aforementioned long-slit observations (Section~\ref{sec:comparison}) at our disposal which we can use to break this degeneracy for the following reason.  In Figure~\ref{fig:R23} we plot the upper and lower branch relations of the $R_{23}$ index against oxygen abundance.  Over-plotted are the metallicities derived from these two relations using pixels from the $R_{23}$ image.  We also show the oxygen abundance calculated by TI05, 12+log(O/H)$=7.89\pm0.1$, using the `direct method' (a highly accurate method that utilises the electron temperature of the gas), which crosses the lower-branch of the $R_{23}$ index.

An offset of approximately $-0.15$~dex exists between the direct-method abundance of TI05 and the mean $R_{23}$ metallicity calculated from our images (shown as a black star).  This offset is not unexpected - direct-method abundances are often found to be slightly lower than those calculated from strong-line indices, which we typically attribute to model uncertainties, temperature fluctuations and/or deviations from thermal electron distributions \citep[e.g. see][and references therein]{Belfiore:2015}.  To check that the discrepancy is not due to our emission-line ratios, we also plot the lower-branch $R_{23}$ metallicities as derived from the emission-line fluxes of TI05 and MK06 (green), which are seen to be in agreement with our distribution along the lower-branch.  Further evidence that the O/H abundance of Mrk~71 lies on the lower-branch of the index can be found in the emission line diagnostic diagram shown in Figures~\ref{fig:BPT_col} and \ref{fig:mod}, where the emission line flux ratios lie in regions typically occupied by low-metallicity objects \citep[see e.g.,][]{Dopita:2013}.

Once convinced that Mrk~71 metallicities are represented by the lower-branch $R_{23}$ index we create a `metallicity image' of Mrk~71, which we show in Figure~\ref{fig:met}.  Whilst the index-based metallicities may not represent the `absolute' metallicity of the gas, we can use this image to investigate the distribution of metallicity throughout Mrk~71 and determine whether or not Mrk~71 is chemically homogeneous.  From all pixels within Figure~\ref{fig:met} we calculate a mean of 12+$\log$(O/H)$=8.04\pm0.11$ (represented by the datapoint within the figure's colour bar).  If we compare this average metallicity to the distribution of metals shown in Figure~\ref{fig:met}, we can see that not only is Mrk~71 chemically \textit{in}-homogeneous, but also that a number of interesting chemical structures can be seen in the O/H dispersal throughout the system.  

Firstly, the gas appears to increase in metallicity with increasing distance from cluster B.  This positive gradient is seen most clearly across the edge of the arm that extends northwards from cluster B, a region $\sim$45~pc across.  The gradient is also mirrored in NGC~2366-II.  Positive, or inverted, metallicity gradients are relatively rare and have so far been attributed to interaction and/or metal-poor gas falling into the centre of the system \citep{Cresci:2010,Queyrel:2012,Sanchez-Almeida:2014b}.   While we are yet to detect inflowing gas towards Mrk~71, \foiii\ and \hi\ maps do suggest that it is interacting with NGC~2363 (not covered here), $\sim$1.4~kpc north-west of Mrk~71 \citep{Roy:1991,Braun:1995}. 

Secondly, there is a ring-shaped decrease in metallicity surrounding cluster A ($\sim$26~pc in diameter and highlighted by a red-dashed box).  We investigate the authenticity of this structure in the lower panels of Fig.~\ref{fig:met}, where we show radial cuts from the centre of the highlighted region (as shown in inner panel).  Each cross-section profile represents the average metallicity in each radial bin along each coloured segment, and incorporates the total uncertainty in 12+log(O/H).  It can be seen that in segments covering the most pronounced part of the ring-shaped structure (i.e. the lower edge of the ring, panel b), the decrease in metallicity is over a $\sim$5--10~pc distance and exists outside the uncertainties in each bin.  However, in less pronounced areas of the ring the gas appears to be chemically homogeneous within the uncertainties.

\begin{figure*}
\includegraphics[scale=0.6]{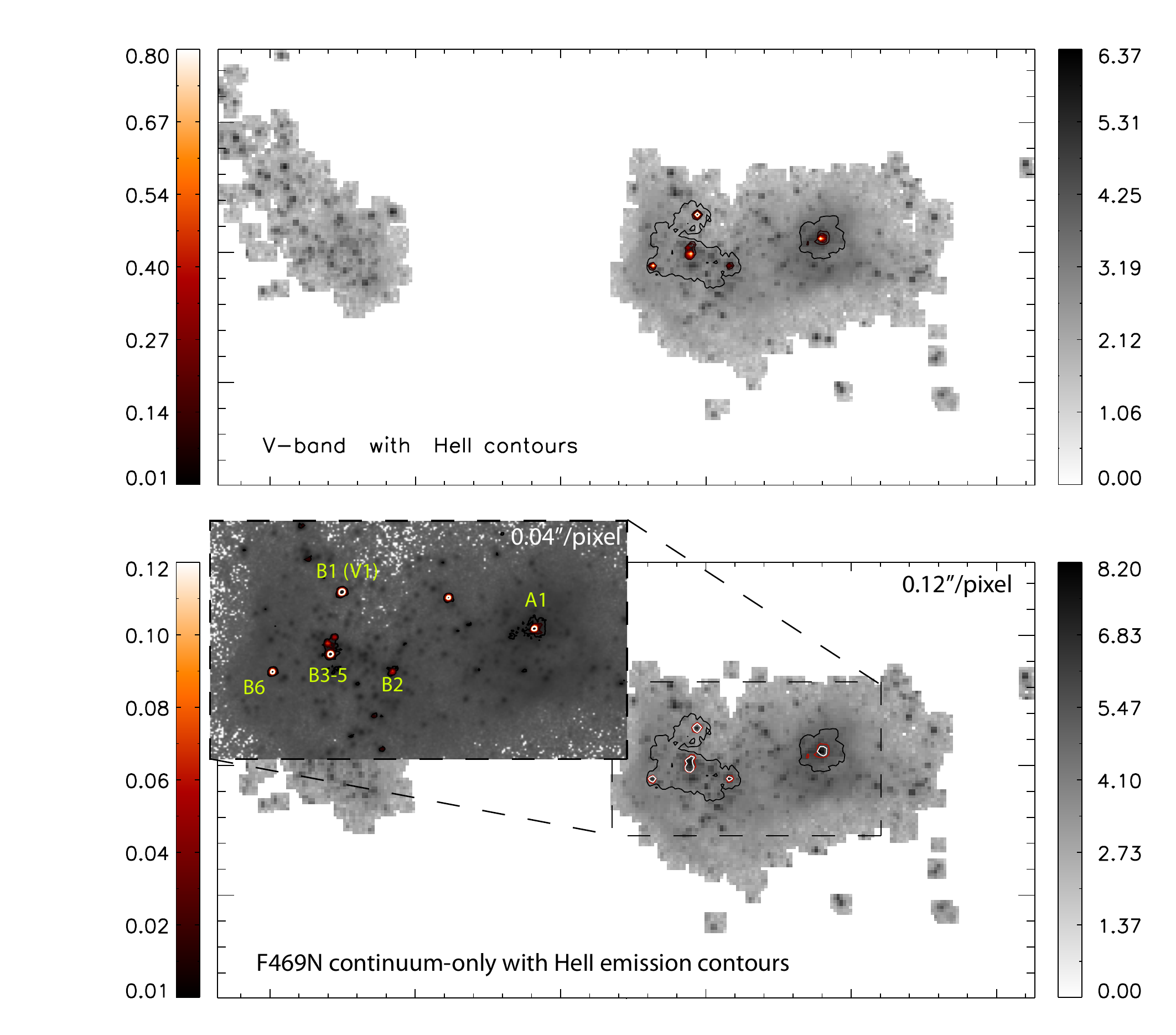}
\caption{A comparison between the morphology of \heii\ emission and $V$-band stellar continuum \textit{(top-panel)}, and F469N continuum-only (\textit{bottom-panel}).  The inset shows the original (0.04$"$/pixel) F469N continuum-only image with \heii\ contours overlaid, which we use to resolve individual WR stars. For each contoured image we mask 50\%\ of the peak flux in order to show the low-level emission.}\label{fig:heii_conts}
\end{figure*}

If we consider the young ($<1$~Myr) age of cluster A, this lower ring of low-metallicity gas may be interpreted as gas that is yet to be enriched by the stellar population within the cluster.  However, this scenario would not explain how the metallicity then increases outside the ring. A similar scenario has also been witnessed in NGC~5253, a local starburst galaxy famed for hosting the first observed case of nitrogen-enhancement \citep[e.g.][]{Walsh:1989,Kobulnicky:1997,Lopez-Sanchez:2007,Westmoquette:2013}.  Like Mrk~71, the chemical inhomogeneity in NGC~5253 lies very close to a region hosting extremely young ($\lesssim1$~Myr) star clusters \citep{Calzetti:2015}.  This suggests that the extreme youth of the clusters may be playing a role here, i.e. the pollution from the most massive stars within the cluster have not yet mixed properly and as a result we observe chemical inhomogeneities.

An alternative (and probably more likely) scenario is that here we are seeing a region in which the  photoionisation-model-based R$_{23}$ diagnostic breaks down. Indeed, the ring of low-metallicity gas aligns well with the ionised-gas `bubble' traced by the strong \foii\ and \fsii\ emission (Fig.~\ref{fig:nbands}), which may signify a shock-front.  However, we know from Section~\ref{sec:BPT} that non-photoionised gas does not exist in this region.   It is perhaps instead due to the fact that strong-line diagnostics are designed to work over \hii\ regions as a whole, i.e. where the emission is integrated over the sphere of gas photoionised by the central stellar source.  As discussed in Section~\ref{sec:results}, the bubble blow-out structure seen in this region is analogous to a single Str{\"o}mgren sphere and metallicity diagnostics such as R$_{23}$ cannot be applied because it was calibrated over whole, or multiple, spheres of this kind.  The diagnostic may not work correctly here because we are instead spatially resolving emission \textit{across} the \hii\ region. 

\subsection{\heii\ emission line imaging}\label{sec:heii}

The narrow-band F469N observations were designed to locate the extent of \heii\,$\lambda$4686 emission throughout Mrk~71.  This emission has two contributing factors.  Firstly, WR stars can produce characteristic broad emission (FWHM $\gtrsim$1000~\kms) in specific ions or `WR-features' (e.g. \heii\,$\lambda$4686) that originate in the envelopes of massive stars undergoing rapid mass loss.  Secondly, a narrow nebular \heii~$\lambda$4686 emission line (FWHM $\sim$100--300~\kms) can be superimposed upon this broad feature.  This line, with an ionisation potential of 4~Ry, is often seen in the spectra of BCDs due to the fact that the hardness of the ionising radiation increases with decreasing metallicity (e.g., Campbell et al. 1986).  The origin of nebular \heii~$\lambda$4686 has often been cause for debate \citep{Thuan:2005,Izotov:2007,Kehrig:2015}, with suggestions including hot WR stars, shocks from supernovae, and X-ray binaries.  Previous long-slit observations of Mrk~71 suggest that both broad and narrow spectral features may be present.  One (out of three) optical spectra of Mrk~71 obtained by TI05, and shown in detail by \citet{Izotov:2007}, shows prominent WR features, whereas nebular \heii\,$\lambda$4686 is observed in all three spectra.   As we are unable to disentangle nebular from broad emission in our \heii\ image, here we can only treat the combined emission as a tracer for highly-ionised gas and WR stars. 
 
The 1$\sigma$ \heii\ (continuum-subtracted) image is shown in Figure~\ref{fig:nbands} (with 3$\sigma$ contours overlaid), where localised emission can be seen in the two main stellar clusters - a strong peak associated with cluster A and another in and surrounding cluster B.  Each peak of \heii\ emission is 0.12--0.16 arcsec across, corresponding to 2--3~pc at the distance of Mrk~71.  In order to explore the source of our \heii\ detections, in Figure~\ref{fig:heii_conts} we show the \heii\ emission overlaid on the 1$\sigma$ cut $V$-band image and F469N continuum-only image (i.e. a by-product of our continuum-fitting process).  Both images show only the underlying stellar continuum, i.e. without WR or nebular \heii\ emission, and it can be seen that each peak in \heii\  emission aligns with stellar continuum features, suggesting that the emission is (at least) partly due to WR stars.

\begin{table}
\caption{\heii\ emission measured within a circular aperture centred on knots of \heii\ emission labelled in Figures~\ref{fig:heii_conts} and~\ref{fig:O3Hb_HeII}.}
\begin{center}
\begin{tabular}{lcc}
\tableline
WR region & Flux / $\times10^{-15} erg s^{-1} cm^{-2}$ & $L$ / $ \times10^{36} erg s^{-1}$\\
\tableline
A1 & 5.88 & 3.01 \\
B1$^\star$ & 3.58 & 1.84\\
B2$^\star$ & 1.44 & 7.38\\
B3-5 & 5.68 & 2.91\\
B6$^\star$ & 2.06 & 1.06\\
\tableline
\end{tabular}
\\[1.5mm]
Notes: Each aperture has a radius of 0.7$''$, apart from region A with a radius of 1.15$''$.\\
$^\star$ Not present at the 3$\sigma$ detection level.
\label{tab:WR}
\end{center}

\end{table}%

\begin{figure}
\includegraphics[scale=0.5]{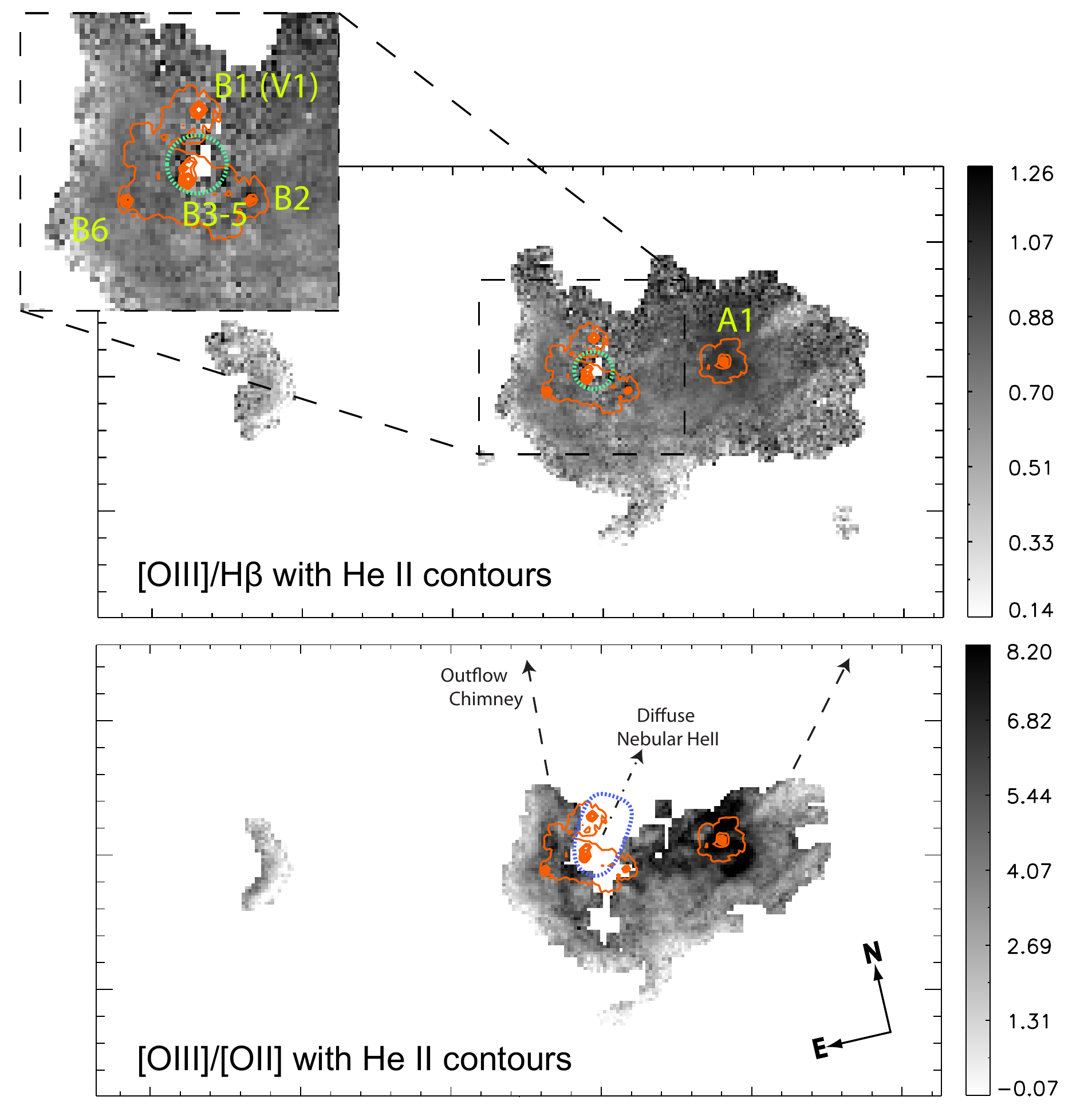}
\caption{The morphology and strength of ionising field throughout Mrk~71, with \heii\  contours overlaid (orange). Knots of \heii\ emission are labelled according to the clusters in which they lie, with corresponding fluxes given in Table~\ref{tab:WR}. \textit{Top-panel:} \foiii/\hb\ image with an inset showing a zoom-in of the area defined by the black dashed box.  This highlights a small blown-out cavity, $\sim$18~pc in diameter (green dotted circle) where no \foiii\ or \hb\ exists.  \textit{Bottom-panel:}  \foiii/\foii, which corresponds roughly to the ionisation parameter, $U$. Here we highlight a larger cavity, $\sim$28~pc across, where no \foii\ emission exists (blue dotted ellipse).  Both cavities align with the direction of extended nebular \heii\ emission seen by \citet{Drissen:2000} (dot-dashed arrow) and the `chimney' of outflowing gas discovered by \citet{Roy:1991} (dashed arrow).}\label{fig:O3Hb_HeII}
\end{figure}

We label each peak of \heii\ emission in Figure~\ref{fig:O3Hb_HeII}, according to the cluster in which it is located. The summed \heii\ flux within a 1.15$''$ radius aperture for A1 and a 0.7$''$ aperture for B1--B6 is listed in Table~\ref{tab:WR}, along with the corresponding luminosities (at a distance of $D=3.44$~Mpc). If we inspect the original, un-binned \heii\ image (i.e. at 0.04$''$/pixel, inset of Figure~\ref{fig:heii_conts}) we can see that each aperture contains a single WR star, with the exception of B3-5 which contains a cluster of three WR stars.  The additional peak in \heii\ emission in the 0.04$''$/pixel image (between the two clusters and unlabelled) is not visible in the $1\sigma$ cut 0.12$''$/pixel image, as such we do not consider it as a reliable detection.  We must also bear in mind that at the 3$\sigma$ level (contours overlaid in Figure~\ref{fig:nbands}), \heii\ emission is seen only B3--5 and A1, suggesting that cluster B may only host three WR stars overall.

Using a similar method (i.e. F469N WFPC2 0.05\arcsec/pixel observations), \citet{Drissen:2000} also located three WR stars in cluster B at the location of B3-5, but no significant \heii\ emission outside B3-5 or in cluster A.  They attribute the latter to the fact that cluster A is too young to harbour a WR population and that the starlight is hidden by dust.  However, this may also in-part be due to an over-subtraction of their continuum image, which was created from a weighted average of F547M and F439W images, with the latter being contaminated by a number of emission lines (Figure~\ref{fig:spec}).     \citet{Drissen:2000} also detect a WR star 150~pc north of cluster B which we do not see in our 0.04$''$/pixel or 0.12$''$/pixel \heii\ images.  

\begin{figure}
\includegraphics[scale=0.5,angle=0]{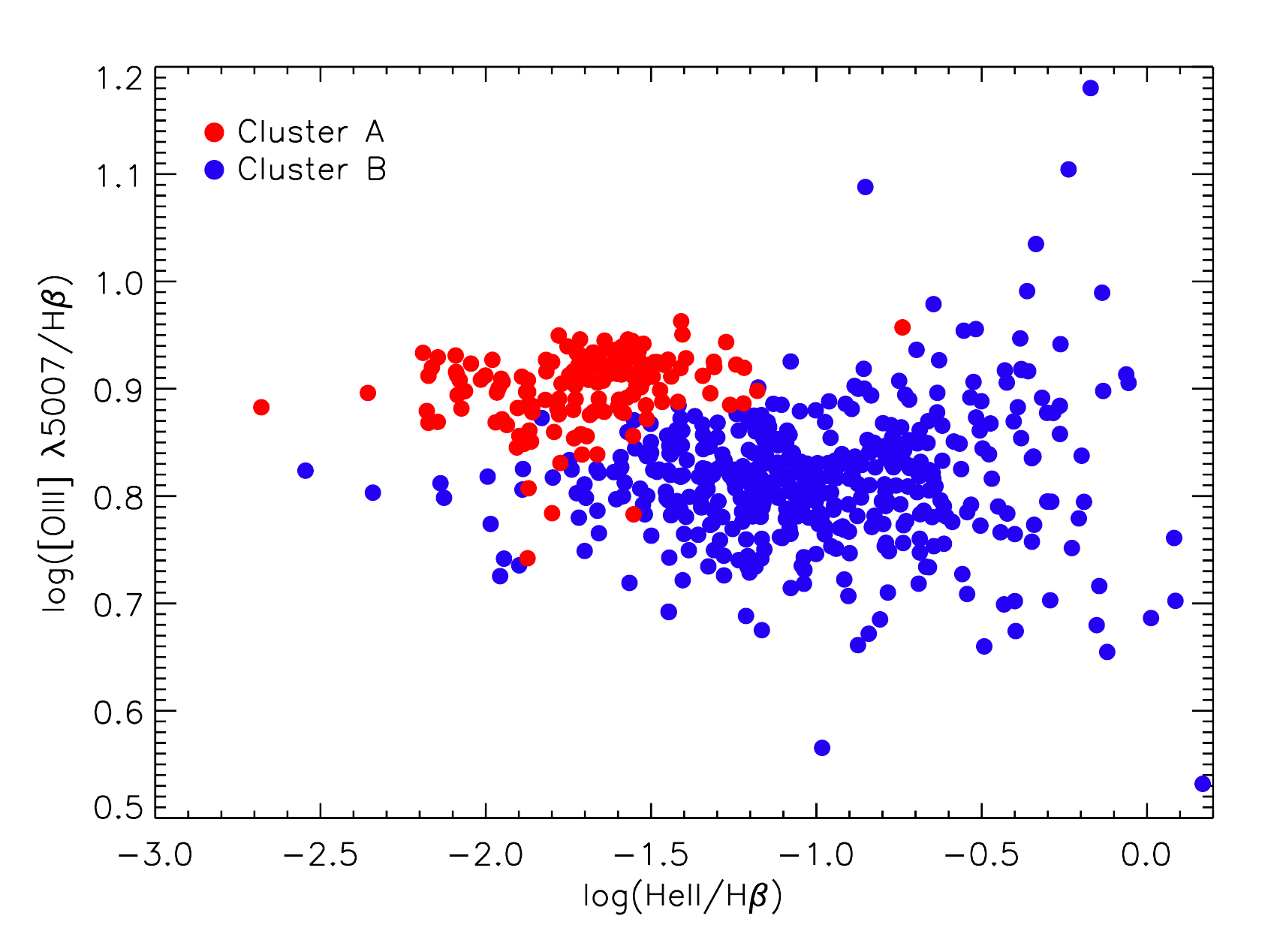}
\caption{The distribution of \heii/\hb\ emission with regards to the degree of ionisation, represented by \foiii/\hb. Cluster A is seen to occupy a higher \foiii/\hb\ ratio than cluster B, suggesting that \heii\ emission in this region is from both nebular+broad emission.}\label{fig:HeII_O3Hb}
\end{figure}

We can investigate the nebular component of \heii\ emission by comparing its morphology to that of the highly-ionised gas in Mrk~71.  The strength of the ionising radiation can be parameterised by \foiii/\foii\, (created from reddening-corrected images) and \foiii/\hb\ emission line ratios, which are shown in Figure~\ref{fig:O3Hb_HeII} with \heii\ contours overlaid.  The most highly-ionised gas is located around cluster A, from which large filaments of strong \foiii/\foii\ are seen to extend. The strongest peak in \heii\ emission aligns with the most strongly ionised gas (i.e. cluster A), whereas the emission along the arm and cluster B lies amongst relatively low-ionisation gas.   Combined with its young age \citep[$\lesssim1$~Myr,][]{Drissen:2000}, this suggests that the \heii\ flux detected in cluster A is most likely nebular in origin, whereas cluster B's emission is attributable to  broad \heii\ emission from WR stars.  To illustrate this further, in Figure~\ref{fig:HeII_O3Hb} we plot \heii/\hb\ verses \foiii/\hb\ for regions of gas surrounding clusters A and B (regions are denoted in Figure~\ref{fig:nbands}).  While no clear correlation exists between \heii\ emission and the strength of the ionising radiation, the two clusters do occupy different regions of this parameter space, with cluster A showing a higher \foiii/\hb\ ratio than cluster B by $\sim$0.15~dex, demonstrating that cluster A does indeed have harder ionising radiation than cluster B.  

WR stars are known to harbour powerful stellar winds, with wind densities an order of magnitude higher than massive O stars, which can have a profound influence on the surrounding ISM \citep{Crowther:2007}. The inset of Figure~\ref{fig:O3Hb_HeII} shows a zoom-in around cluster B where there is also a region ($\sim18$~pc across, green-dashed circle) of little or no \foiii/\hb\ flux surrounding the brightest knot of \heii\ emission.  The lack of emission in this region is suggestive of a blow-out cavity possibly caused by the winds from the B3-5 WR cluster.  The cavity has a much larger extent for \foii\ (28~pc across, blue-dashed ellipse in lower panel of Fig.~\ref{fig:O3Hb_HeII}) than \foiii\ and \hb \footnote{This `gap' in emission line flux also exists in the $1\sigma$ and $2\sigma$ cut \foii, \foiii, and \hb\ images, demonstrating that it is not simply due to lack of signal in this region.}, and aligns well with the direction of extended nebular \heii\ emission seen by \citet{Drissen:2000} (their figure 8) and the north-northwest orientation of the large `chimney' of outflowing gas discovered by \citet{Roy:1991}. It should also be noted that the WR stars in cluster B are also joined by dozens of OB stars \citep[Figure~\ref{fig:colour}, see also][]{Drissen:2000} which would contribute to the energy ejected into the ISM here, along with supernova explosions.  \citet{Drissen:2000} estimate that the total energy returned to the ISM via winds from the stars in the core of cluster B, over 2.5~Myr since the starburst, would be $5\times10^{50}$~erg.  While this is two orders of magnitude less than the kinetic energy in the expanding super-bubble surrounding Mrk~71 \citep{Roy:1991,Roy:1992}, the stars within the core of cluster B may be capable of creating the small cavity in the surrounding ISM that we see here.

\section{Discussion}

The observations presented here provide insight into a local star-forming galaxy on 2~pc scales.  What do we gain from observing star-forming galaxies on such fine spatial scales?  The first, and perhaps most obvious gain, is that we can observe complex structure in the ionising gas, both with respect to the morphology of the gas and also the physical feedback effects of stars (bubbles, blow-out etc).  For example, we have seen that a cluster of three WR stars may be capable of blowing out a region $\sim$18--28~pc in size.   Using the morphology of emission lines we can investigate the presence of thin filamentary fronts (typically $\sim$10~pc), changes in the ionisation state of the gas, and also possible chemical structure.

The fine spatial scales utilised here enable us to `image' rather than `map' the physical conditions throughout Mrk~71.  To illustrate this concept, in Figure~\ref{fig:feedback} we show colour-composite images constructed from images of metallicity ($R_{23}$), ionisation strength (\foiii/\foii), ionised gas (\ha\ emission), and hot, young stars ($U$-band image).  Each colour channel uses the original (0.04$''$/pixel) narrow-band images and are not masked with respect to the signal in each pixel.  Despite the lack of quantitative value, these images can be used to investigate the effects of feedback throughout Mrk~71.  For example, in the top-panel of Fig.~\ref{fig:feedback} the \ha\ emission (red) is seen to dominate along edges surrounding the highly-ionised gas (green), with metallicity gradient peaking along the outer edges of the galaxy.  The edge of the ionised bubble, which may have caused the ring of low-metallicity gas shown in Fig.~\ref{fig:met}, is clearly apparent, and the decrease in ionisation from the inside of the bubble is seen as a ring of yellow.  In the bottom panel, $R_{23}$ is replaced with the $U$-band, which traces the continuum emission of hot, young stars.  While cluster B can be seen clearly (in the `crux' of the arm), cluster A is embedded behind the \ha\ emission on the north-east edge of the \ha\ emission.  Both pictures are in line with the findings of \citet{Drissen:2000}, who suggest that cluster A is so young ($<1$~Myr) that it is still embedded in dust, although a significant fraction of the UV radiation emitted by its hidden massive stars manages to leak out of the dust to provide the bulk of ionising flux for Mrk~71. In NGC~2366-II there is a scattering of stars along the eastern edge - most likely the cause of the ionised gas front and \fsii/\ha\ and \foiii/\hb\ gradients seen in Figures~\ref{fig:BPT_col} and \ref{fig:mod}.

Despite the large amounts of energy being injected into the ISM throughout Mrk~71, our emission line diagnostic images clearly show that the gas is predominantly photoionised. However, the lack of detectable shock-excitation remains puzzling given the young, intense star-formation and consequently outflowing gas.  This non-detection may in-part be due to the fact that our observations lack kinematical information. If we had instead observed Mrk~71 with a high spatial resolution IFU we would undoubtably see emission-lines with multiple velocity components along the line of sight, thereby providing a far more complete and accurate view of the ISM.  For example, gas moving at different velocities along the line of sight would give a range of line centroids and gas at different temperatures would produce profiles with different line widths (due to line broadening).  Of course, we would still be unable to disentangle velocity components perpendicular to the line of sight, so a favourable geometry would also be beneficial here.  Nevertheless, using individual line kinematics, one could de-compose the emission line components to obtain separate and more accurate line ratios, a technique commonly used in IFU studies \citep{James:2009,James:2013a, James:2013b,Amorin:2012, Hagele:2012, Westmoquette:2013}.  As demonstrated by \citet{Rich:2011, Rich:2014}, this method can be especially effective in isolating shock-excited emission which may at present lost amongst the more dominant emission-line components from photoionisation.

We find the \heii\ emission from cluster A rather intriguing, given the extreme youth of the cluster ($\lesssim$1~Myr).  Its age dictates that the emission cannot be due to WR stars and, despite the fact that the \heii/\hb\ flux ratio from cluster A can be reconciled via photoionisation models, it is only done so by assuming input spectra from planetary nebula nuclei which would also exist long after the cluster's age.  One way to achieve nebular \heii\ emission in a young stellar system would be from the hard ionising flux from Very Massive Stars (VMS) \citep{Crowther:2010}.  Such stars, which can be as massive as 300~\Msol, may be present within young clusters and would have powerful stellar winds and a significantly increased ionising flux compared to lower mass stars.  As such, both broad and narrow \heii\ emission features would be present and reminiscent of WR populations, despite the clusters being at an early evolutionary phase \citep[1--2~Myr,][]{Crowther:2010}.  The existence of VMS has also been proposed in NGC~5253 by \citet{Calzetti:2015} to explain the ionising flux from the extremely young clusters ($\lesssim$1~Myr) located within the `radio nebula', where nitrogen-rich WR stars have also been detected.

\begin{figure}
\includegraphics[scale=0.8,angle=0]{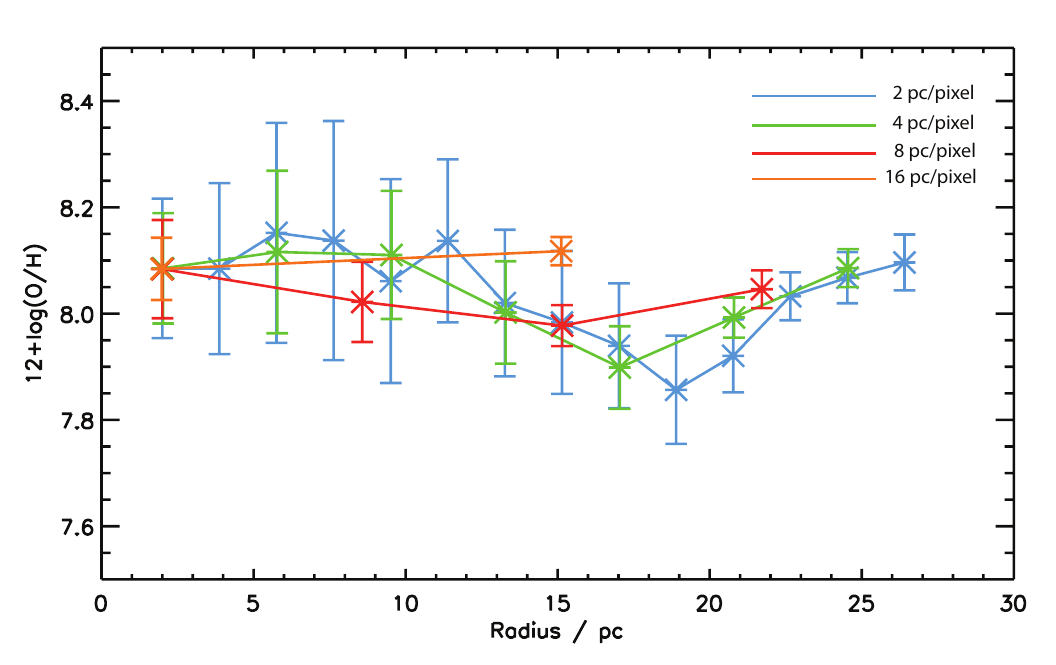}
\caption{The effects of decreasing spatial resolution.  Here we show the radial gradient originally shown in Figure~\ref{fig:met}b at 2~pc/pixel resolution, degraded to 4, 8 and 16~pc/pixel resolution.}\label{fig:gradbins}
\end{figure}

The detailed chemical structure observed in Mrk~71 (which we discuss in Section~\ref{sec:met}) would of course be lost if we instead imaged Mrk~71 on larger spatial scales.  This is illustrated in Figure~\ref{fig:gradbins}, where we show the radial gradient originally shown in Figure~\ref{fig:met}b at 2~pc/pixel resolution, degraded to 4, 8 and 16~pc/pixel resolution, and the gradient is seen to diminish almost entirely at 8~pc/pixel.  While unsurprising, this loss in structure with resolution does highlight the need for care when assessing metallicity gradients and their evolution across cosmic time.  As shown by \citet{Yuan:2013}, for each individual metallicity gradient there is a critical angular resolution FWHM below which the measured metallicity gradient is flattened.  This has a significant consequence when interpreting the apparent steepening of metallicity gradients with redshift, especially when the highest redshift observations are aided by gravitational lensing and adaptive optics.   However, it should be noted that the spatial resolution does \textit{not} affect the calculation of the galaxy's average metallicity - from all pixels within Figure~\ref{fig:met} we calculate a mean of 12+log(O/H)=8.04$\pm$0.11 and if we use the total, reddening corrected fluxes from the \foiii, \foii\ and \hb\ images and the lower-branch $R_{23}$ diagnostic, we derive 12+log(O/H)=8.02$\pm$0.15.  This effect was first demonstrated by \citet{Kobulnicky:1998} and more recently by \citet{Mast:2014}, where IFU observations of spiral galaxies at $\sim$45~pc/$''$ are simulated up to 1~kpc/$''$ to show that any smooth, global trends remain observable at larger spatial scales but any sharp structure is lost.  These results provide hope for the use of metallicity diagnostics on high-$z$ galaxies where spatially resolved line ratios are not available and we are required to integrate light over the entire galaxy.

While the data presented here clearly illustrates the gains from high-spatial resolution observations, it also demonstrates the need for care when working with emission-line diagnostics on such scales.  As discussed previously in Section~\ref{sec:met}, emission-line-based metallicity calibrations have been designed to provide us with information integrated over entire \hii\ regions rather than \textit{throughout} \hii\ regions.  In addition to this, emission line diagnostics were derived from photoionisation models and, at $2$~pc scales, excitation due to photoionisation may not dominate. Both of these points need to be taken into consideration as we enter the era of extremely large telescopes.  At the end of the next decade, observing gas structure on such small scales will not only be limited to the local Universe with both ground and space-based observatories, with telescopes such as the E-ELT, GMT, and TMT being designed to probe $<50$~pc scales out to redshifts as high as $z=$2.  Despite engaging with the fine-scale ISM structure that we have presented here, we may still need to integrate over $>20$~pc regions if we intend to continue using strong-line ratio diagnostics.

\begin{figure*}
\includegraphics[scale=0.7]{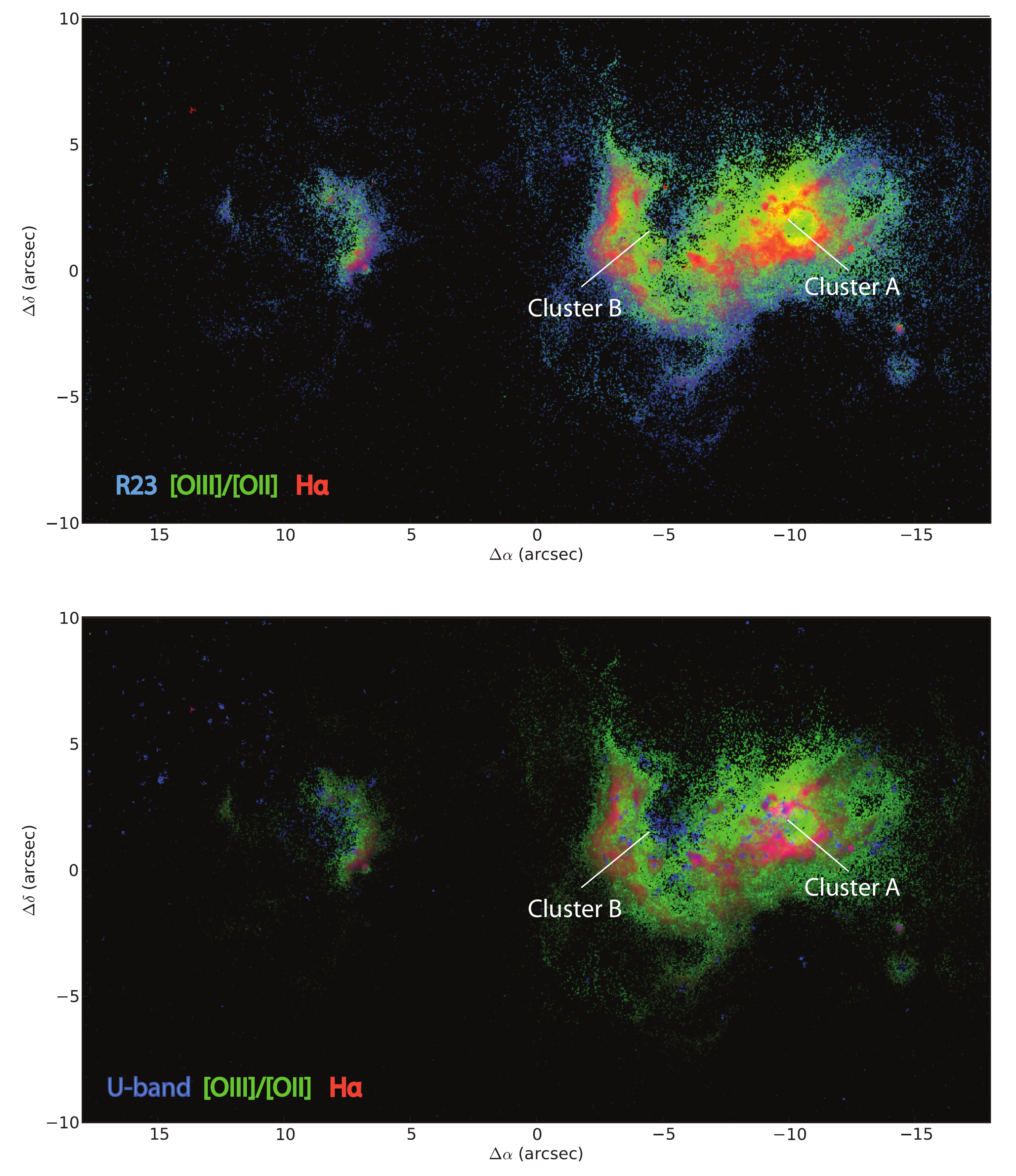}
\caption{To illustrate the benefits of ultra-high resolution emission-line imaging, here we create a colour composite of Mrk~71 in $R_{23}$ (i.e. metallicity, blue), \foiii/\foii\ (i.e. ionisation strength, green) and \ha\ (ionised gas, red). In the bottom-panel we replace $R_{23}$ with the U-band, to highlight the position of hot, young stars.  The location of the two main clusters (A and B) within Mrk~71 are labelled in both images.}\label{fig:feedback}
\end{figure*}

\section{Conclusions}

The paper presents an extensive dataset of new \textit{HST}-WFC3 observations of Mrk~71, one component of a nearby ($D\sim3.44$~Mpc) blue-compact dwarf galaxy containing one of the most powerful local starbursts known.  The data consist of high-resolution ($0.04''$/pixel) images in seven narrow-band and four broad-band filters, enabling us to create continuum-subtracted images of Mrk~71 in the light of \foii~($\lambda3727+\lambda3729$), \heii~$\lambda$4686, \hb, \foiii~$\lambda$5007, \ha, and \fsii~($\lambda6716+\lambda6731$).  The final binned data provide emission line images for all lines at spatial resolutions of $\sim$2~pc/pixel at $3\sigma$ confidence levels, with the exception of \heii\ which we use at the $1\sigma$ confidence level.  NGC~2366-II is also seen within our data, although it is not the primary focus of our work. 

The data presented here complement a plethora of multi-wavelength observations of Mrk~71, which have been documented extensively in the literature.  From these studies, we know Mrk~71 to be a low-metallicity ($\sim0.17$~\Zsol) giant \hii\ region in NGC~2366 which consists of two main stellar clusters, A and B (Figure~\ref{fig:colour}).  It is known to harbour an expanding super-bubble which envelopes the entire system and a chimney of outflowing gas extending north-northwest away from the super-bubble.  Mrk~71 is perhaps most famous for displaying signs of hyper-sonic gas (lines with $FWHM\sim2400$~\kms) around cluster A.

Using the high-resolution emission line images we explore the chemical and physical properties of the ionised gas throughout Mrk~71 in unprecedented detail and deduce the following:

\begin{itemize}[leftmargin=0cm]
\item Each emission line image shows one clear peak in flux surrounding cluster A, a super star cluster, whereas cluster B is located in a cavity of emission with an arm of ionised gas extending northwards along its eastern edge.  The morphology of the ionised gas is less extended in the lower ionisation lines, \fsii\ and \foii, than in the \foiii\ and Balmer line images.  Both the \fsii\ and \foii\ images show a pronounced shell of emission around cluster A, which we interpret as the edges of an ionised gas bubble $\sim$22~pc in diameter or single \hii\ region.

\item The \foiii/\hb\ and \fsii/\ha\ line ratio images show an extensive amount of structure, with a strong peak in \foiii/\hb\ surrounding cluster A, and a distinct gradient in both ratios as you move away from cluster A and along the edges of cluster B and NGC2366-II.  Putting both ratios onto an emission line diagnostic plot on a pixel-by-pixel basis we find the gas to be predominantly photoionised throughout, with no sign of shock excitation.  We confirm the findings of \citet{Drissen:2000} who suggest that cluster A is responsible for the bulk of ionising photons for the galaxy, although our line ratio images suggest evidence for a non-negligible contribution from cluster B also.  Photoionisation models reveal the gas to be of high density, with $\log(U)=-1$ to $-2$, between 0.15--0.2~\Zsol, and a range of opacities.

\item Using the \foiii, \foii\, and \hb\ images and the strong-line metallicity diagnostic, $R_{23}$, we create the first `metallicity image' of a galaxy, i.e. on spatial scales of 2~pc.  The image reveals that the gas within Mrk~71 is not chemically homogeneous, with chemical structure on scales larger than the 1$\sigma$ distribution of metallicity throughout the system.  There is an increase in metallicity outwards from cluster B and across the eastern arm ($\sim$45~pc in width).  Moreover, a ring-shaped decrease in metallicity is seen surrounding cluster A, $\sim$26~pc across with an edge $\sim$5--10~pc thick, which aligns with the shell of \foii\ and \fsii\ emission discussed above.  We attribute this structure to the fact that we are spatially resolving gas across a Str{\"o}mgren sphere rather than integrating over it, and here the $R_{23}$ diagnostic, which was designed to work over entire \hii\ regions, breaks down.

\item The \heii\ image, which traces both the 4650~\AA\ WR feature and the nebular \heii~$\lambda$4686 emission, reveals five strong peaks in emission, one in cluster A and four in cluster B.  Upon inspection of the un-binned \heii\ image (i.e. at 0.04$''$/pixel) the peaks were resolved into individual WR stars, with the strongest peak in cluster B showing a cluster of three WR stars.   However, at the 3$\sigma$ level, only the strongest peaks in clusters A and B are detected. Considering the young age of cluster A and its high density of ionising photons, we conclude that the \heii\ emission within this cluster is either purely nebular in origin or due to very massive stars. We therefore detect a total of three WR stars in Mrk~71 (each located in cluster B, although three more may potentially be present).

\item A small blow-out cavity is detected at the edge of the WR cluster in cluster B, which is $\sim$18~pc across in \foiii\ and \hb, and $\sim$28~pc across in \foii\ emission.  The cavities align well with the direction of extended nebular \heii\ emission seen by \citet{Drissen:2000} and the `chimney' of outflowing gas \citep{Roy:1991}.  We suspect that the energy ejected into the ISM from this WR cluster, combined with that from dozens of massive OB stars seen in the vicinity, may be responsible for blowing out the ionised gas surrounding cluster B and creating this emission-line cavity.

\item We demonstrate that emission line images of this kind can be used in their ratioed form to create `feedback images', i.e. colour-composites that show the physical conditions of the gas, such as the strength of the ionising radiation (\foiii/\oii) or metallicity ($R_{23}$), along with the ionising sources themselves.  Such images can be used to investigate the effects of feedback throughout star-forming systems.  For Mrk~71, these images highlight the structure of the ionised gas in relation to the ionising sources, the extent of the  \hii\ region surrounding cluster A  and the ionisation gradient along its edge.
\end{itemize}

The observations presented here offer an insight into the ultra-high spatial resolution observations achievable from both current and future ground-based AO-assisted IFU instruments. High spatial-resolution emission line images have enabled us to explore the chemical and physical structure of the ISM on a parsec-by-parsec basis throughout a low-metallicity star-forming galaxy - a local analogue to young galaxies in the high-$z$ universe. By observing galaxies of this kind on such fine spatial scales, we can observe the full complexity of the ionised gas, both with regards to its morphology and ionisation state, and also the physical effects of the stellar feedback (e.g. outflows/superwinds) in primordial-like systems.  In Mrk~71, the energy injected into the ISM from the winds of massive stars has succeeded in blowing out a large cavity of ionised gas, possibly contributing to a large-scale outflow of gas away from the main body.  Regulating mechanisms such as this are thought to suppress star-formation and prevent low mass halos from forming dwarf galaxies.  On the other hand, stellar superwinds are providing an efficient mechanism for transporting photons and material outside of the dwarf galaxy, suggesting that small systems of this kind may play a role in the re-ionisation of the Universe at $z=$6--11.

Our metallicity images enable us to observe chemical inhomogeneity on $<50$~pc scales and (apparent) changes in chemical structure across Mrk~71 on scales as small as 2~pc/pixel.  Although the structure is lost at degraded resolutions, it does not affect the average measured metallicity of the galaxy, offering hope for the use of metallicity diagnostics on high-$z$ galaxies where light is integrated over large spatial scales.

While the gains of high-resolution emission line images are obvious, we have also shown that the use of strong-line diagnostics may not be possible in this regime.  For example, metallicity diagnostics were designed and calibrated on emission line ratios integrated over entire \hii\ regions and using them on spatially resolved emission across a \hii\ region may cause erroneous or misleading results.  Such limitations on our much-loved strong-line methods are especially important as we enter the era of 30--40~m telescopes and start probing $<50$~pc scales at $z=2$.

Finally we note that the data presented here are inherently flawed in comparison to the future high spatial-resolution ($<0.2''$) IFU spectroscopic observations, which will allow us to access the third dimension of our emission-line maps - velocity.  High resolution velocity maps of the outflowing gas from cluster B and the ionised bubble surrounding cluster A, would enable us to explore the full complexity of feedback effects, e.g. by mapping expansion velocities and mass-loading factors for the gas entrained in each structure, and uncovering shock excitation from separated emission line components.  

Observations such as those presented here provide excellent benchmarks for the high spatial resolution hydrodynamical simulations that endeavour to model a `realistic' ISM, with an aim to understanding the triggering and evolution of star-formation in galaxies throughout the universe.

\bibliographystyle{mn2e}
\bibliography{references.bib}

\acknowledgments
The authors give thanks to Roberto Maiolino,  Max Pettini, and Francesco Belfiore for invaluable discussions concerning the content of this paper, and to Rob Kennicutt for his insightful comments on this manuscript.  We also greatly appreciated discussions and assistance from  Mike Dopita and Ralph Sutherland regarding photoionisation modelling with \textsc{mappings}.  We would like to sincerely thank Sungryong Hong for assisting us in the testing continuum-subtraction methods and Jay Anderson for help with WFC3 image reduction.  We sincerely thank the reviewer of the paper whose helpful comments and suggestions
greatly improved the paper.  STScI is operated by the Association of Universities for Research in Astronomy, Inc., under NASA contract NAS5-26555.  Support for Program number 13041 was provided by NASA through a grant from the Space Telescope Science Institute, which is op-erated by the Association of Universities for Research in Astronomy, Incorporated, under NASA contract NAS5-26555.

{\it Facilities:} \facility{HST (WFC3)}

\clearpage

\clearpage

\end{document}